\documentclass[aps,preprint,graphicx,groupedaddress,amsmath,amssymb]{revtex4-2}

\usepackage{subcaption, graphicx}
\usepackage{color, float}
\usepackage{bm}
\usepackage{hyperref}
\usepackage{eucal, listings}
\begin{document}

\title{Out-of-equilibrium modeling of lyotropic liquid crystals: from binary simulations to multi-component theory}


\author{Jonathan Salmerón-Hernández}
\email{Contact author: jsalmeron@uchicago.edu}
\thanks{Current Address: Tandon School of Engineering, New York University, Brooklyn, NY 11201, USA }
\affiliation{Pritzker School of Molecular Engineering, University of Chicago, Chicago, IL 60637}

\author{Pablo Zubieta-Rico}
\affiliation{Pritzker School of Molecular Engineering, University of Chicago, Chicago, IL 60637}

\author{Juan de Pablo}
\email{Contact author: jjd8110@nyu.edu}
\affiliation{Tandon School of Engineering, New York University, Brooklyn, NY 11201, USA}
\affiliation{Pritzker School of Molecular Engineering, University of Chicago, Chicago, IL 60637}

\date{\today}

\begin{abstract}

We present a thermodynamically-consistent theoretical framework for lyotropic liquid crystals (LCs) that is based on the GENERIC (General Equation for the Non-Equilibrium Reversible-Irreversible Coupling) formalism. This formalism ensures conservation of energy and production of entropy, while coupling concentration, momentum balance, and liquid crystalline order. Starting from a binary nematic–isotropic mixture, we derive a theory for these key variables, which is then extended to multi-component systems.

The binary equations are solved numerically using a Julia-based solver that relies on an upwind finite-difference scheme, thereby enabling stable and efficient simulations that are capable of handling multiple time scales while satisfying fundamental mathematical constraints. The results of simulations are shown to be consistent with experimental observations of topological core defects in chromonic LCs, as well as flow-driven droplet shape transitions under Couette and Poiseuille flows.

This work provides a platform for simulations of multi-component lyotropic LCs that can be extended to systems with multiple interfaces, active materials, and materials subject to external fields.
\end{abstract}

\maketitle

\section{Introduction}

Liquid crystalline materials are encountered in a wide range of settings, from biological systems to advanced technologies. Most of those settings involve multiple components and interfaces, and generally function under conditions that are far from equilibrium. It is widely appreciated that the coupling of orientational ordering to mass and momentum transport can lead to emergent phenomena such as phase transitions, self-organization, and defect formation \cite{deGennes1993physics, NarayanSwarming2007, taveravazquez2023lightactivated, Majumdar2014PerspectivesIA}. Understanding how such phenomena arise is crucial for design of materials with tailored properties, ranging from complex fluids and polymer blends to tissue-like composite systems \cite{Kumareaat7779, RanftFluidization, LCLiver2019}. 

Lyotropic liquid crystals (LCs) represent a particular class of mixtures that exhibit partially ordered states whose behavior is largely driven by concentration gradients. These mixtures should be contrasted to thermotropic uni-component LCs, where phase transitions are primarily driven by temperature changes. Recent work on lyotropic LCs, often involving an active nematic in contact with an isotropic fluid \cite{wilking2011biofilms, boyer2011buckling, darnton2004moving}, has revealed a range of complex behaviors involving the coupling of flow and structure that remain unexplained \cite{wu2017transition, guillamat2016control}. Models capable of describing such systems are in their infancy \cite{Sulaimanetal06, carenza2019lattice, liu2019nucleation, assante2023active}; importantly, the vast majority of theoretical and computational work published on liquid crystalline materials to date has been focused on single component uni-phase systems at equilibrium.

Early computational efforts, particularly those employing thermotropic LC-based models with the Beris–Edwards (BE) formalism \cite{BerisEdwards}, have generated valuable insights into lyotropic systems \cite{Kumareaat7779, head2024spontaneous}. These studies have helped explain a range of new phenomena, such as activity-induced defect patterns, elastic bands, and droplet deformation \cite{zhang2021spatiotemporal, zhang2018interplay, zhang2016controlled}. However, treating concentration explicitly as a dynamical variable in simulations continues to be a challenge, thus placing limits on our understanding of, for example, local effects such as concentration changes near defect cores \cite{wilking2011biofilms, Kumareaat7779, zhou2017fine}.

In what follows we present a theoretical framework for multicomponent, multiphasic liquid crystalline systems at equilibrium and far from equilibrium. Our approach is based on the GENERIC (General Equation for the Non-Equilibrium Reversible-Irreversible Coupling), thereby leading to a thermodynamically consistent model for such materials, including lyotropic. The dynamics of out-of-equilibrium systems is governed by a time-evolution equation of the form \cite{ottinger2005beyond, PavelkaKlikaGrmela, grmela1997dynamics, ottinger1997dynamics}:
 \begin{equation} \label{eq:1}
 \frac{\partial \bm{x}}{\partial t} = \bm{L}(\bm{x})\cdot\frac{\delta E(\bm{x})}{\delta \bm{x}} + \bm{M}(\bm{x}) \cdot \frac{\delta S(\bm{x})}{\delta \bm{x}},
 \end{equation}
where $\bm{x}$ represents a set of field variables, such as density and momentum, and $t$ denotes time. The boldface notation refers to tensorial variables. The linear operators, $\bm{L}$ and $\bm{M}$, also known as Poisson and friction matrices, play a crucial role in coupling the relevant physical mechanisms under consideration (e.g. convection and diffusion). The total energy ($E$) and the entropy ($S$) are real-valued functionals written in terms of $\bm{x}$, while the ($\cdot$) symbol denotes dot product. Additionally, $\delta_{\bm{x}}$ refers to a functional derivative. In essence, Eq. (\ref{eq:1}) expresses the evolution of any system as the sum of reversible and irreversible contributions: the energy functional derivative corresponds to reversible phenomena, where the total energy is conserved and the system can in principle return to its initial state, while the second term is associated with irreversible processes, which produce entropy.

The GENERIC equation (\ref{eq:1}) not only reproduces several fundamental equations in physics, such as the Navier-Stokes and Maxwell equations\cite{ottinger1997dynamics, jelic2006dissipative}, but also, at its core, embodies a synthesis of thermodynamics. Importantly, GENERIC is compatible with the Poisson bracket formalism \cite{Edwards+1998+301+333}, or the theoretical methodology utilized by BE to formulate their approach.

Our manuscript is organized as follows. Sec. \ref{Model} summarizes the GENERIC building blocks: the total energy and entropy, along with the Poisson and friction matrices. These are intended to describe, as an initial step, a two-component nematic-isotropic mixture. A generalization of this theory to multi-component systems is provided in Appendix \ref{Multicomponent}. Section \ref{Results} discusses the considerations employed to solve the underlying equations and simulate experimentally-inspired systems. The results obtained with our own open-source library, written in the Julia programming language, are compared to experimental observations.

More specifically, we showcase three different scenarios, each with increasing complexity. First, we consider a chromonic LC systems with $\pm\,1/2$ topological defects, for which we examine the steady-state morphology. Second, we explore the evolution configuration of axial droplets under two different flow conditions, Couette and Poisseulle (parabolic). In section \ref{Conclusions} we summarize our findings.

\section{The Model}\label{Model}

We construct our model by defining the dynamical variables $\bm{x}$, followed by the total energy and entropy densities, as well as their corresponding Poisson and friction matrices. The hydrodynamics are then dictated by Eq. (\ref{eq:1}).

\subsection{Variables}

Variables $\bm{x}$, encompass the total mass density $\rho$, the total momentum density $\bm{m}$, and the internal energy density $\epsilon$—the three classical hydrodynamic fields for simple fluids, which have also been widely adopted in the GENERIC formalism \cite{DEPABLO2001137}. The composition or mass fraction of the liquid crystal is given by $\phi$ ($0 \le \phi \le 1$), hereafter referred to as the LC concentration, and the second-order tensor $\bm{Q}$ describes the liquid crystalline order. Note that the concentration of the second component, an isotropic liquid, can be expressed as $1-\phi$. In short, the vector $\bm{x} = ( \rho, \bm{m}, \epsilon, \phi, \bm{Q})$ comprises five different position-dependent fields ($\bm{r}$).

We choose the $\bm{Q}$ tensor as it is well-defined even near singularities, such as topological defects \cite{deGennes1993physics}. In this work, we focus on the uniaxial case given by:
\begin{equation}\label{Qdef}
{Q}_{ij} = q\left(n_i n_j - \frac{{I}_{ij}}{d} \,  \right),
\end{equation}
where $n_i$ represents the unit director vector, which corresponds to the average orientation of the LC, and satisfies the head-to-tail symmetry ($\bm{n}\equiv-\bm{n}$). The scalar order parameter $q$ ($0 \le q \le 1$) is $0$ for an isotropic state and $1$ for perfect alignment. $I_{ij}$ denotes the identity matrix and $d$ indicates the dimensionality of the system (e.g., 2D or 3D). By construction, $\bm{Q}$ is symmetric ($Q_{ij} = Q_{ji}$) and traceless ($Q_{ll} = 0$), which helps to enforce the constraints in $q$ and $\bm{n}$. Here, the Latin subscripts (i, j, k, l, m) indicate the basis components, and we assume the Einstein summation convention over repeated indices (unless otherwise stated). 

For visualization purposes, we present our simulation results in terms of $q$ and $\bm{n}$, which correspond to the highest eigenvalue of $\bm{Q}$ and its associated eigenvector \cite{ball2010q}. Note that while we consider the uniaxial approximation for simplicity in our calculations, our model can readily be extended to materials that exhibit significant biaxiality without major modifications. However, this extension requires computing an additional order parameter and a secondary director vector \cite{liu2019nucleation}.

\subsection{Energy} \label{Energy}

The total energy $E(\bm{x})$ is defined by integrating the total energy density $\mathcal{E}(\bm{x})$ over the entire constant volume ($V$) of the system
 \begin{eqnarray} \label{totalE}
E(\bm{x}) &=& \int \mathcal{E}(\bm{x}) \, dV  =  \int \bigg[
 \frac{1}{2} \frac{{m_{k}}^{2} }{\rho} + \epsilon
 + \frac{1}{2} \kappa_E \! \left(\frac{\partial\phi}{\partial r_k}\right)^2 + \frac{1}{2} \kappa' \!
 \left(\frac{\partial\phi}{\partial r_i}\right) \left( Q_{ij} + \frac{I_{ij}}{d}  \right) \left(\frac{\partial\phi}{\partial r_j}\right)
\nonumber \\ &+& \frac{1}{2} K(\phi)  \frac{\partial Q_{ij}}{\partial r_k} \frac{\partial Q_{ij}}{\partial r_k} \, + \, \rho \chi \phi (1-\phi)
 + \epsilon_{\rm LdG,E}( \rho, \phi, \bm{Q}) 
 \bigg] dV ,
 \end{eqnarray}

this energy density, which overall promotes ordering, includes the kinetic, internal, interfacial, and  anchoring energies, as well as elastic Frank-Oseen (FO), Flory-Huggins (FH), and Landau-de Gennes (LdG) contributions.

The first and second terms represent the total kinetic energy density and $\epsilon$ within the integral. The third term incorporates diffuse interfaces into our model, often used in the phase field and LC literature \cite{Sulaimanetal06, deGennes1980dynamics}. This latter term defines the phase interface as the region with the largest variation in $\phi$. Here, such variations are smooth or diffuse; they arise from spatial derivatives that ensure a continuous concentration function and prevent non-differentiable sharp changes. The coefficient $\kappa_E$ controls the size or thickness of this interfacial region \cite{cahn1958free}.

Note that a diffuse interface approach is adopted for simplicity and ease of numerical implementation \cite{provatas2011phase}, and we recognize its limitations for computing interfacial properties \cite{rauscher2022nonequilibrium}. These limitations stem from the assumption that bulk thermodynamic relationships can be readily applied to interfacial variables. This is not always valid, especially at equilibrium (see \cite{rowlinson2013molecular}), and imposes limits on the spatial resolution of the interfacial thickness. To address these issues, one could adopt a sharp interface description \cite{provatas2011phase}, through GENERIC and small system thermodynamics  \cite{hill1994thermodynamics}, and treat the interface as an additional independent dynamical variable.

The anchoring energy density (fourth term) serves to couple the LC orientation with the concentration gradient \cite{araki2004nematohydrodynamic}. Here, we introduce an additional isotropic term $(\bm{I}/d)$ \cite{hashiguchi2020nonlinear} (this should be contrasted with Eq. (9) of \cite{Sulaimanetal06}). We restrict our model to one coupling term since the effect of $\kappa'$ on the orientation of the director vector at the interface is well understood (see below). 

The fifth term represents the classical FO energy density, under the approximation of a single elastic constant $(K)$, which penalizes gradients of $ \bm{Q}$ and captures the orientational elasticity on the length scale of the liquid crystal \cite{frank1958liquid}.
The coefficients $\kappa_E$, $\kappa'$ and $K$ have dimensions of [force], or [energy/length]; $\kappa_E$ and $K$ are expected to be positive, while $\kappa'$ can be of either sign, depending on whether the preferred orientation of the liquid crystal is perpendicular ($\kappa' < 0$) or parallel ($\kappa' > 0$) to the interface \cite{Sulaimanetal06}. 
  
The second-to-last term denotes the FH energy density, describing the repulsive interaction between the two liquids \cite{flory1953principles}, with its positive $\chi$ having dimensions of [energy/mass]. The last term $\epsilon_{\rm LdG,E}$ controls the isotropic-nematic transition. To align with the GENERIC framework, we split the conventional LdG free energy into its energetic and entropic contributions, following the classical analysis of Landau \cite{sheng1974landau}. Here, we introduce only its energetic part, represented by the quartic form \cite{doi1981molecular}
\begin{equation}\label{ELdG}
\epsilon_{\rm LdG,E}( \rho, \phi, \bm{Q}) = - \rho B(\phi) \left[
\frac{1}{6} \, Q_{ij}Q_{ji}
+ \frac{1}{3} \, Q_{ij}Q_{jk}Q_{ki}
- \frac{1}{4} \, (Q_{ij}Q_{ji})^2
\right] ,
\end{equation}
where parameter $B(\phi)$ has dimensions of [energy/mass]. Additionally, $B(\phi)$ should be proportional to $\phi^2$, as $\epsilon_{\rm LdG,E}$ arises from pairwise interactions between LC units \cite{doi1981molecular,Sulaimanetal06}. The overall factor $\rho$ gives the contributions the character of an extensive density.

Moving forward, we take functional derivatives with respect to the field variables $\bm{x}$, as depicted in Eq. (\ref{eq:1}), to obtain
\begin{equation}\label{Egrad}
\frac{\delta E}{\delta \bm{x}} =
\left( \!\!\!
\begin{array}{cc}
\frac{\delta E}{\delta  \rho} \\
\frac{\delta E}{\delta  \bm{m}} \\
\frac{\delta E}{\delta  \epsilon} \\
\frac{\delta E}{\delta  \phi} \\ \,\,\,\,
\frac{\delta E}{\delta   \bm{Q}}^\mathtt{ST}
\end{array}
\! \right)  = 
\left(
\begin{array}{c}
- \frac{1}{2} {v_{k}}^2 + \chi \phi (1-\phi) + \frac{\epsilon_{\rm LdG,E}}{\rho}\\
v_{k} \\
1 \\
\mu_1
 + \mu_2 
+ \mu_3 \\
\bm{H}_{an}
+\bm{H}_{el}
+\bm{H}_{LdG}
\end{array}
\right) ,
\end{equation}
where
\begin{equation}\label{mu_1}
    \mu_1 = - \kappa_E 
 \frac{\partial}{\partial r_k}\frac{\partial \phi}{\partial {r_k}} 
 - \kappa' \frac{\partial}{\partial r_k}
 \!\!\left[\left( Q_{kl}
 + \frac{I_{kl}}{d}\right) 
  \frac{\partial \phi}{\partial r_l} \right] ,
\end{equation}
\begin{equation}\label{mu_2}
    \mu_2 = \frac{1}{2}\frac{\partial K}{\partial\phi} \left( \frac{\partial Q_{ij}}{\partial r_k} \right)^2 ,
\end{equation}
\begin{equation}\label{mu_3}
    \mu_3 = \rho \chi (1-2\phi) + \frac{1}{B} \frac{\partial B}{\partial \phi} \, \epsilon_{\rm LdG,E} ,
\end{equation}
\begin{equation}\label{H_an}
    (H_{an})_{ij} = \frac{1}{2} \kappa' \! \left[ \left(\frac{\partial \phi}{\partial r_i}\right) \left(\frac{\partial \phi}{\partial r_j}\right)
-   \frac{I_{ij}}{d}   \left(\frac{\partial \phi}{\partial r_l}\right)^2  \, \right],
\end{equation}
\begin{equation}\label{H_el}
    (H_{el})_{ij} = - K \frac{\partial}{\partial r_k}\frac{\partial Q_{ij}}{\partial {r_k}}   \, ,
\end{equation}
\begin{equation}\label{H_LdG}
    (H_{LdG})_{ij} = - \rho B(\phi) \left[ \frac{1}{3}Q_{ij} + Q_{ik}Q_{kj} -\left( Q_{ij} + \frac{I_{ij}}{d}\right) Q_{lk}Q_{kl}  \right] \, ,
\end{equation}
and we further introduce the velocity field $\bm{v}=\bm{m}/\rho$. In accordance with standard conventions, we denote the expressions appearing in the derivatives of $\phi$ and $\bm{Q}$ with $\mu$ and $\bm{H}$, commonly related to the chemical potential and molecular field, respectively. As usual, the symmetric and traceless $(\mathtt{ST})$ properties of $\bm{Q}$ are also reflected in its associated functional derivative
\begin{equation}\label{ST_derivative}
   \left( \frac{\delta E}{\delta Q_{ij}} \right)^{\mathtt{ST}} =
   \frac{\delta E}{\delta Q_{ij}} - \frac{I_{ij}}{d} \left( \frac{\delta E}{\delta \bm{Q}} \right)_{ll}, 
\end{equation}
where the second term on the right-hand side (RHS) acts as a Lagrange multiplier, ensuring the traceless condition, and adds terms proportional to $\bm{I}/d$ in equations (\ref{H_an}) and (\ref{H_LdG}). Hereafter, we omit the $(\mathtt{ST})$ superscripts when referring to the functional derivatives of $\bm{Q}$ to simplify notation.

\subsection{Entropy} \label{Entropy}

The entropy $S$ is also written as an integral over $V$. We divide this entropy into two terms: 1) the background fluid entropy density $s_0(\rho, \epsilon)$, used to later define the background pressure and the system temperature, and 2) any additional entropic contribution $\mathcal{S}(\bm{x})$, 
\begin{align} \label{totalS}
S(\bm{x}) &= \int  \bigg[s_0(\rho, \epsilon) + \mathcal{S}(\bm{x})  \bigg] dV 
\nonumber \\ &= \int \left[
s_0( \rho, \epsilon) + \frac{1}{2} \kappa_S \! \left(\frac{\partial\phi}{\partial r_k}\right)^2
+ s_{\rm mix}( \rho, \phi) + 
\epsilon_{\rm LdG,S}( \rho, \phi, \bm{Q})
\right] dV.
\end{align}

Here, $\mathcal{S}$ encompasses the interfacial and mixing entropy densities, as well as the LdG entropic effect, which  collectively favor disorder. We incorporate the $(\partial \phi / \partial r_k)^2$ term in equations (\ref{totalE}) and (\ref{totalS}), since its resulting interfacial tension could contain both energetic and entropic elements.
The constant parameter $\kappa_S$ has dimensions of [entropy/length] or [force/temperature], with its subscript $S$ denoting its entropic effect. Correspondingly, in Eq. (\ref{totalE}), the subscript $E$ of $\kappa_E$  denotes an energetic contribution.

The entropy of mixing, which counteracts the FH energetic term, takes the conventional form
\begin{equation}\label{Smix}
s_{\rm mix}( \rho, \phi) = - \rho \Big[  C_1 \, \phi \ln \phi
+ C_2 (1-\phi) \ln (1-\phi) \Big] ,
\end{equation}
consistent with the literature on mixtures \cite{atkins2023atkins}. Here, the coefficients $C_1$ ($C_2$) represent Boltzmann's constant divided by the mass of one liquid crystal (or isotropic liquid) molecule. This term confines $\phi$ to the physical range $(0,1)$, which cannot be enforced by the frequently used quartic  free-energy expression $(1-\phi^2)^2$ \cite{Sulaimanetal06, carenza2019lattice, cahn1958free}.

The last term in Eq. (\ref{totalS}), or the entropic part of the LdG free energy, balances the effect of  $\epsilon_{\rm LdG,E}$, and reads
\begin{equation} \label{ELdGS}
\epsilon_{\rm LdG,S}( \rho, \phi, \bm{Q}) =  - \frac{1}{2} \rho A(\phi)  \, Q_{ij}Q_{ji} ,
\end{equation}
 where the dimensions of $A$ are [entropy/mass]. A common practice in the LC literature \cite{Sulaimanetal06} introduces a factor of $\phi^2$ for both functions $A(\phi)$ and $B(\phi)$. While we understand the rationale behind  $B(\phi)$ (stemming from pair-wise interactions), this reasoning might not extend to the entropic contribution $A(\phi)$, which is associated with Brownian forces and expected to be proportional to $\phi$. The ratio of $B(\phi)/A(\phi) \propto \phi$ aligns well with the theory developed by \cite{doi1981molecular}. A more explicit form for both $A(\phi)$ and $B(\phi) $ is introduced below when solving the resulting time-evolution equations. 

Finally, we compute the derivatives of the entropy functional, as dictated in Eq. (\ref{eq:1}), to arrive at

\begin{equation}\label{Sgrad}
\frac{\delta S}{\delta \bm{x}} =  \left(
\begin{array}{c}
\frac{\partial s_0}{\partial \rho} + \frac{s_{\rm mix}}{\rho}
- \frac{1}{2} A(\phi) Q_{ij}Q_{ji} \\
0_k \\
1/T \\ 
- \kappa_S \frac{\partial}{\partial r_k}\frac{\partial \phi}{\partial {r_k}} 
- \rho \Big[  C_1 \ln \phi - C_2 \ln (1-\phi) + C_1 - C_2 \Big]
- \frac{1}{2} \frac{\partial A}{\partial \phi} \, \rho \,Q_{ij}Q_{ji} \\
- A(\phi) \rho \, Q_{ij}
\end{array}
\right),
\end{equation}
where the third component of this gradient allows us to define the background fluid temperature $T$ by $1/T=\partial s_0/\partial \epsilon$.

\subsection{Poisson Operator}

Based on equations (2.60) and (4.44) of \cite{ottinger2005beyond} and (11.5-37) of \cite{BerisEdwards}, we introduce the following $\bm{L}$ matrix, or Poisson Operator, which encodes how variables interact and affect each other in a reversible and dynamic manner,

\begin{equation}\label{LCPoisson}
\bm{L} = - \left(
\begin{array}{ccccc}
0 & \frac{\partial}{\partial r_k} \rho & 0 & 0 & 0 \\
\rho \frac{\partial}{\partial r_i}
& \frac{\partial}{\partial r_k} m_i + m_k \frac{\partial}{\partial r_i}
& \epsilon \frac{\partial}{\partial r_i} + \frac{\partial}{\partial r_m} \pi^{S}_{mi}
& - \frac{\partial \phi}{\partial r_i} & \bm{L}_{25} \\
0 & \frac{\partial}{\partial r_k} \epsilon + \pi^{S }_{km} \frac{\partial}{\partial r_m}
& 0 & 0 & 0 \\
0 & \frac{\partial \phi}{\partial r_k} & 0 & 0 & 0 \\
0 & \bm{L}_{52} & 0 & 0 & 0 \\
\end{array} \right) ,
\end{equation}
with
\begin{eqnarray}
\bm{L}_{25} &=& - \frac{\partial Q_{kl}}{\partial r_i}
- \frac{\partial}{\partial r_m} \left[
I_{ki} \left( Q_{ml} + \frac{I_{ml}}{d}   \right)
+ \left( Q_{km} + \frac{I_{km}}{d}  \right) I_{il} \right]
\nonumber \\
&+& 2 \frac{\partial}{\partial r_m} \left( Q_{mi} + \frac{I_{mi}}{d}  \right)
\left( Q_{kl} + \frac{I_{kl}}{d}  \right) ,
\label{L25expr}
\end{eqnarray}
and
\begin{eqnarray}
\bm{L}_{52} &=& \frac{\partial Q_{ij}}{\partial r_k}
- \left[ I_{ik} \left( Q_{mj} + \frac{I_{mj}}{d}   \right)
+ \left( Q_{im} + \frac{I_{im}}{d}  \right) I_{kj}
\right] \frac{\partial}{\partial r_m}
\nonumber \\
&+& 2 \left( Q_{ij} + \frac{I_{ij}}{d}  \right)
\left( Q_{km} + \frac{I_{km}}{d}  \right) \frac{\partial}{\partial r_m} ,
\label{L52expr}
\end{eqnarray}
where the indices $i,j$ and $k,l$ refer to the left side (rows) and right side (columns), respectively.

In Eq. (\ref{LCPoisson}) and onward, we adopt GENERIC conventions, where derivatives of the form $\frac{\partial \psi}{\partial r_m}$ act \emph{only} on $\psi$, an auxiliary function here, while $\frac{\partial}{\partial r_m}$ applies differentiation to all subsequent terms, particularly after matrix multiplication. We also clarify that the rows and columns in the GENERIC matrices follow the order of $\bm{x}$. For instance, the (2,4)-element represents terms associated with the second and fourth variables ($\bm{m}$ and $\phi$) and can be interpreted as the reversible response of $\bm{m}$ induced by $\phi$. 

 Since most of the terms shown in Eq. (\ref{LCPoisson}) have been extensively described in \cite{ottinger2005beyond} and \cite{ottinger1997dynamics}, we only summarize the key ideas behind its construction. First, when convection—or the idealized transport of a quantity in a frictionless medium—is the only reversible effect considered in the model, all non-zero entries appear in the entries associated with $\bm{m}$, i.e., the fields respond only through momentum. Furthermore, the $\bm{L}$ matrix must also be antisymmetric, a restriction that ensures energy conservation. Here, antisymmetry refers to a flip in signs or in the ordering of operators. Specifically, the (1,2)-element gives rise to the continuity equation upon multiplication with Eq. (\ref{Egrad}), while the (2,1)-element results from the antisymmetry condition. 
 
Second, hydrodynamic variables have well-established entries in $\bm{L}$, based on their tensorial orders and physical properties \cite{ottinger2005beyond, PavelkaKlikaGrmela}. With this in mind, the (2,2)-element is chosen for the "lower-convected" vector density $\bm{m}$, and the (2,4)- and (4,2)-elements are used to describe scalars, such as $\phi$. Additionally, the (2,5)- and (5,2)-elements are derived from (11.5-37) of \cite{BerisEdwards}, the LC $\mathbf{Q}$-tensor Poisson operator proposed by BE in integral form (see Appendix \ref{BE:appendix} for the transformation from the integral to matrix form). Lastly, the (2,3)- and (3,2)-elements arise when describing the scalar density $\epsilon$, which has a spatial derivative similar to that of other scalar densities, such as $\rho$, but also includes the pressure tensor's entropic contribution $\bm{\pi}^S$.

To understand the physical interpretation of $\bm{\pi}^S$, note that in simple fluids, volume changes, such as compression, affect not only the pressure but also the \textit{internal energy}, which, in turn, impacts the momentum balance. These volume changes are typically described by velocity gradients through a Newtonian stress tensor. However, in complex fluids, the (scalar) pressure and the Newtonian tensor alone cannot capture the effects of the fluid's internal structure. To account for these, an additional pressure tensor contribution is therefore required in the entries associated with $\epsilon$ and $\bm{m}$. Mathematically, this tensor is determined by the first GENERIC degeneracy condition:
\begin{equation} \label{eq:deg1}
\bm{L}(\bm{x})\cdot \frac{\delta S(\bm{x})}{\delta \bm{x}} = 0,
\end{equation} 
which ensures that reversible processes do not produce entropy. 

The resulting entropic tensor, whose detailed derivation is found in the Appendix \ref{appendix:b}, reads as follows:
\begin{eqnarray}\label{entstressfull}
     \pi_{mi}^S &=& T  \bigg\{
\left[ \rho \frac{\partial}{\partial r_k} \left( \frac{\partial \mathcal{S}}{\partial (\partial_k \rho)} \right)  + 
 \epsilon \frac{\partial}{\partial r_k} \left( \frac{\partial \mathcal{S}}{\partial (\partial_k \epsilon)} \right)
- \rho \frac{\partial \mathcal{S}}{\partial \rho} - \epsilon \frac{\partial \mathcal{S}}{\partial \epsilon} + \mathcal{S} + \frac{p_0}{T}  \right]  I_{mi}  
\nonumber \\
&-& \bigg[ \left( \frac{\partial \rho}{\partial r_i} \right) \left(\frac{\partial \mathcal{S}}{\partial (\partial_m \rho)} \right)  
+ \left( \frac{\partial \epsilon}{\partial r_i} \right) \left(\frac{\partial \mathcal{S}}{\partial (\partial_m \epsilon)} \right)    \nonumber \\
&+& \left( \frac{\partial \phi}{\partial r_i} \right) \left(\frac{\partial \mathcal{S}}{\partial (\partial_m \phi)} \right) 
+ \left( \frac{\partial Q_{kl}}{\partial r_i} \right) \left(\frac{\partial \mathcal{S}}{\partial (\partial_m Q_{kl})} \right)   \bigg]  \nonumber \\
&-&  2 \bigg[ \left(Q_{mi} + \frac{I_{mi}}{d}\right) \left(Q_{kl} + \frac{I_{kl}}{d}\right) \frac{\delta S}{\delta Q_{kl}} - \left( Q_{ml} + \frac{I_{ml}}{d} \right) \frac{\delta S}{\delta Q_{li}}  \bigg]
  \bigg\}  .
\end{eqnarray}

Here, the background pressure $p_0$ is given by the local-equilibrium Euler equation,
\begin{equation}\label{p}
\frac{p_0}{T} = s_0 - \rho \frac{\partial s_0}{\partial \rho}
- \epsilon \frac{\partial s_0}{\partial \epsilon},
\end{equation}

and $T$, as defined in Eq. (\ref{Sgrad}), appears naturally in the final form.

\subsection{Friction Matrix}

To construct the friction (or dissipative) matrix $\bm{M}$, which couples irreversible dynamics, we first set the entries associated with $\rho$ to zero. In accordance with GENERIC, we assume that $\bm{M}$ is the sum of multiple matrices, each representing a specific physical mechanism. These matrices must be symmetric, positive semi-definite, and satisfy the second degeneracy condition:
\begin{equation} \label{eq:deg2}
\bm{M}(\bm{x})\cdot \frac{\delta E(\bm{x})}{\delta \bm{x}} = 0.
\end{equation}
These mathematical restrictions ensure that entropy always increases over time \cite{ottinger2005beyond}. With this structure, we choose to incorporate diffusion, relaxation, and slip processes into our model and write:
\begin{equation}\label{Mcont}
\bm{M} =   \bm{M}^{\rm diff}_{\phi}+ \bm{M}^{\rm relax}_{\bm{Q}}  + \bm{M}^{\rm slip}_{\bm{Q}},
\end{equation}
with
\begin{equation}\label{Mdif}
\bm{M}^{\rm diff}_{\phi} = \left( \begin{matrix}

- \frac{\delta E}{\delta \phi} \frac{1}{\rho} \frac{\partial}{\partial \bm{r}}
\cdot \rho\bm{D} \cdot \frac{\partial}{\partial \bm{r}} \frac{1}{\rho} \frac{\delta E}{\delta \phi} 
&  \frac{\delta E}{\delta \phi} \frac{1}{\rho} \frac{\partial}{\partial \bm{r}}
\cdot \rho\bm{D} \cdot \frac{\partial}{\partial \bm{r}} \frac{1}{\rho} \quad
& 0 \\ 
 \frac{1}{\rho} \frac{\partial}{\partial \bm{r}} \cdot \rho\bm{D} \cdot
\frac{\partial}{\partial \bm{r}} \frac{1}{\rho} \frac{\delta E}{\delta \phi} 
& - \frac{1}{\rho} \frac{\partial}{\partial \bm{r}} \cdot \rho\bm{D} \cdot \frac{\partial}{\partial \bm{r}} \frac{1}{\rho} & 0 \\
0 & 0 & 0

\end{matrix} \right) ,
\end{equation}
and
\begin{equation}\label{MrelQ}
\bm{M}^{\rm relax}_{\bm{Q}} = \left( \begin{matrix}
 \frac{\delta E}{\delta \bm{Q}}:\bm{R}:\frac{\delta E}{\delta \bm{Q}}  & 0 & -  \frac{\delta E}{\delta \bm{Q}}:\bm{R} \\
0 & 0 & 0\\
- \mathbf{R}:\frac{\delta E}{\delta \mathbf{Q}}  & 0 & \mathbf{R}

\end{matrix} \right) .
\end{equation}

Here, both $\bm{M}^{\rm diff}_{\phi}$ and $\bm{M}^{\rm relax}_{\bm{Q}}$ represent the 3x3 block of the friction matrix associated with components $\epsilon$, $\phi$ and $\bm{Q}$, while the other entries are zero.  $\bm{M}^{\rm diff}_{\phi}$ describes a diffusion process for the conserved quantity $\phi$, similar to Eq.~(4.65) of \cite{ottinger2005beyond}, while $\bm{M}^{\rm relax}_{\bm{Q}}$ models a relaxation (or minimization) mechanism for the non-conserved variable \(\bm{Q}\), following Eq.~(4.54) of \cite{ottinger2005beyond}.

To describe diffusion processes, we choose to add spatial derivatives, similar to other diffusion's law, like Fick's. The positive semi-definite second-order tensor $\bm{D} = \bm{D}(\rho, \epsilon, \phi, \bm{Q})$ corresponds to anisotropic diffusion. In contrast, because a relaxation process implies the absence of transport, the occurrence of spatial derivatives is not needed, with only a relaxation parameter $\bm{R}$ to drive the system towards a minimum.

A concrete form for the symmetric and positive semi-definite fourth-rank tensor $\bm{R}$ has been proposed in Eq.~(11.5-40) of \cite{BerisEdwards},
\begin{equation}\label{Rexample}
R_{ijkl} = R( \rho,\epsilon, \phi, \bm{Q})
\left[ \frac{1}{2} (I_{ik} \, I_{jl} + I_{il} \, I_{jk}) \right]  ,
\end{equation}
where we drop the $3 Q_{ij} Q_{kl}$ term, in the original equation, to recover the Ginzburg-Landau equation in the absence of flow (e.g., we are working in a non-inertial framework) with a positive relaxation parameter $R( \rho,\epsilon, \phi, \bm{Q})$ that has dimensions of [volume/(time$\cdot$entropy)]. The work of Doi \cite{doi1981molecular} suggests that $R \propto \phi^{-3}$.

Briefly, to determine the non-vanishing entries in $\bm{M}^{\rm relax}_Q$, we proceed as follows: 1) place the tensor $\bm{R}$ in the $\bm{M}^{\rm relax}_Q$ (3,3)-element, corresponding to the driving force $\frac{\delta S}{\delta \bm{Q}}$; 2) satisfy the degeneracy condition (Eq.~\ref{eq:deg2}) in the third row with the (3,1)-element; 3) complete the (1,3)-element by means of the symmetry condition; and 4) fulfill again the degeneracy requirement in the first row with the (1,1)-element. Similarly, we determine the (1,1)-, (1,2)-, and (2,1)-elements of the matrix $\bm{M}^{\rm diff}_{\phi}$ by first placing the spatial derivatives in its center and then applying an analogous procedure.

Next, we introduce the slip matrix $\bm{M}^{\rm slip}_{\bm{Q}}$, which incorporates the slip coefficient $\lambda$. This accounts for the response of $\bm{Q}$ to velocity gradients, leading to a more realistic viscoelastic behavior \cite{gordon1972anisotropic}. This matrix integrates Schowalter derivatives into our model and allows us to recover the BE equation. We base the construction of this matrix on the BE dissipative operator (Eqs.11.4-38 and 11.5-41 of \cite{BerisEdwards}) in its integral form. To translate this into matrix form, we proceed according to the approach outlined in Section V.B of \cite{ottinger1997dynamics}, incorporating ideas from Eq. (4.79) of \cite{ottinger2005beyond} and from Appendix \ref{BE:appendix}; we also include the symmetry and degeneracy conditions discussed above, to complete all non-zero entries. The final ingredient of our model is the resulting $\bm{M}^{\rm slip}_{\bm{Q}}$, which represents the block associated with $\bm{m}$, $\epsilon$, and $\bm{Q}$, and is given by:
\begin{equation}
    \label{slip_matrix}
   \bm{M}^{\rm slip}_{\bm{Q}} =  \left( \begin{matrix}
        0 & \bm{M}_{12} & \bm{M}_{13}\\
        \bm{M}_{21} & 0 & \bm{M}_{23}\\
        \bm{M}_{31} & \bm{M}_{32} & 0
    \end{matrix} \right),
\end{equation}
with
\begin{equation}\label{M_12}
    (M_{12})_{i} =  \frac{\partial}{\partial r_m} \, L_{mikl}^{Q} \, \frac{\delta E}{\delta Q_{kl}} \, T - \frac{\partial}{\partial r_m} \, L_{mijj}^{Q} \left( Q_{kl} + \frac{I_{kl}}{d} \right) \frac{\delta E}{\delta Q_{kl}} \, T , 
\end{equation}
\begin{equation}\label{M_13}
    (M_{13})_{ikl} = - \frac{\partial}{\partial r_m} \, L_{mikl}^{Q} \, T + \frac{\partial}{\partial r_m} \, L_{mijj}^{Q} \left( Q_{kl} + \frac{I_{kl}}{d} \right) \, T,
\end{equation}
\begin{equation}\label{M_21}
    (M_{21})_{k} = T \, \frac{\delta E}{\delta Q_{ij}} \, L_{ijkl}^{Q} \frac{\partial}{\partial r_l} \,    - T \frac{\delta E}{\delta Q_{ij}} \, \left( Q_{ij} + \frac{I_{ij}}{d} \right)  \, L_{kmll}^{Q} \, \frac{\partial}{\partial r_m}  ,
\end{equation}
\begin{equation}\label{M_23}
    (M_{23})_{ij} =- T \, \left[\frac{\partial}{\partial r_k} \frac{\delta E}{\delta m_{l}} \right]  \, L_{klij}^{Q}  \, + T \, \left[ \frac{\partial}{\partial r_k} \frac{\delta E}{\delta m_{l}} \right]  \, \left( Q_{lk} + \frac{I_{lk}}{d} \right)  \, L_{ijmm}^{Q} \,  ,
\end{equation}
\begin{equation}\label{M_31}
    (M_{31})_{mij} =- T \, L_{mijk}^{Q} \frac{\partial}{\partial r_k}  + T \, L_{mill}^{Q}  \, \left( Q_{kj} + \frac{I_{kj}}{d} \right) \frac{\partial}{\partial r_k}  , 
\end{equation}
\begin{equation}\label{M_32}
    (M_{32})_{mi} = T \, L_{mijk}^{Q} \left[\frac{\partial}{\partial r_k} \frac{\delta E}{\delta m_{j}} \right]  \,   \, - T \, L_{mill}^{Q} \left( Q_{kj} + \frac{I_{kj}}{d} \right) \left[ \frac{\partial}{\partial r_k} \frac{\delta E}{\delta m_{j}} \right]  \,   \, \,  ,
\end{equation}
and 
\begin{equation} \label{L_ijkl}
L_{ijkl}^{Q} = \frac{\left( \lambda -1 \right)}{2}  \left[I_{il} \left(Q_{jk} + \frac{I_{jk}}{d}\right) + I_{jl} \left(Q_{ik} + \frac{I_{ik}}{d}\right) 
+ I_{ik} \left(Q_{jl} + \frac{I_{jl}}{d}\right) + I_{jk} \left(Q_{il} + \frac{I_{il}}{d}\right)\right] .
\end{equation}

\subsection{Time-Evolution Equations}\label{PEDSystem}

By inserting all the building blocks—Eqs.~(\ref{Egrad}), (\ref{Sgrad}), (\ref{LCPoisson}), and (\ref{Mcont})—into the GENERIC equation (\ref{eq:1}), we obtain our final time-evolution equations. They recover the BE formulation, now include a dynamic concentration, modify their momentum balance, and introduce a novel equation for the evolution of $\epsilon$. Since these equations are expressed in terms of energy and entropy rather than free energy, as in the BE original formulation, we refer to them as generalized versions. We begin with the continuity equation, which takes its familiar form
\begin{equation}\label{evoleqrho}
\frac{\partial\rho}{\partial t} = - \frac{\partial}{\partial r_k} (\rho \, v_k)  .
\end{equation}

For the momentum balance, or Generalized Navier-Stokes equation, several contributions to the pressure tensor appear
\begin{equation}\label{evoleqM}
\frac{\partial m_i}{\partial t} = -\frac{\partial}{\partial r_m} \bigg[v_m m_i + \pi_{mi}^{S} + \pi_{mi}^E - \pi_{mi}^{slip} \bigg] ,
\end{equation}

where the energetic contribution, $\bm{\pi}^E$, arises from the second component of the matrix multiplication of $\bm{L}(\bm{x}) \cdot \frac{\delta E(\bm{x})}{\delta \bm{x}}$. Its calculation mirrors that of $\bm{\pi}^S$ (Eq.~\ref{entstressfull}), leading to:

\begin{eqnarray}\label{energystressfull}
\pi_{mi}^E &=&  -\, \bigg\{
\left[ \rho \frac{\partial}{\partial r_k} \left( \frac{\partial \mathcal{E}}{\partial (\partial_k \rho)} \right)  + 
\epsilon \frac{\partial}{\partial r_k} \left( \frac{\partial \mathcal{E}}{\partial (\partial_k \epsilon)} \right)
- \rho \frac{\partial \mathcal{E}}{\partial \rho} - \epsilon \frac{\partial \mathcal{E}}{\partial \epsilon} + \mathcal{E}  \right]  I_{mi}  
\nonumber \\
&-& \bigg[ \left( \frac{\partial \rho}{\partial r_i} \right) \left(\frac{\partial \mathcal{E}}{\partial (\partial_m \rho)} \right)  
+ \left( \frac{\partial \epsilon}{\partial r_i} \right) \left(\frac{\partial \mathcal{E}}{\partial (\partial_m \epsilon)} \right)   \nonumber \\
&+& \left( \frac{\partial \phi}{\partial r_i} \right) \left(\frac{\partial \mathcal{E}}{\partial (\partial_m \phi)} \right)
+ \left( \frac{\partial Q_{kl}}{\partial r_i} \right) \left(\frac{\partial \mathcal{E}}{\partial (\partial_m Q_{kl})} \right)   \bigg]  \nonumber \\
&-& 2  \bigg[ \left(Q_{mi} + \frac{I_{mi}}{d}\right) \left(Q_{kl} + \frac{I_{kl}}{d}\right) \frac{\delta E}{\delta Q_{kl}} - \left( Q_{ml} + \frac{I_{ml}}{d} \right) \frac{\delta E}{\delta Q_{li}} \bigg]
\bigg\} .
\end{eqnarray}

Additionally, $\pi_{mi}^{slip}$ is the contribution from $\bm{M}^{\rm slip}_{\bm{Q}}$, specifically from $\bm{M}_{12} \, \frac{\delta S}{\delta \epsilon} + \bm{M}_{13} \, \frac{\delta S}{\delta \bm{Q}}$, and reads
\
\begin{eqnarray}\label{slip_stress}
\pi_{mi}^{slip} &=& \bigg[ \left( Q_{ml} + \frac{I_{ml}}{d} \right) \frac{\delta E}{\delta Q_{li}} + \left( Q_{il} + \frac{I_{il}}{d} \right) \frac{\delta E}{\delta Q_{lm}}  \\ \nonumber
&-&2 \left( Q_{mi} + \frac{I_{mi}}{d}\right)\left( Q_{kl} + \frac{I_{kl}}{d}\right)\frac{\delta E}{\delta Q_{kl}} 
    \\ \nonumber
    &-& \left( Q_{ml} + \frac{I_{ml}}{d} \right) \frac{\delta S} {\delta Q_{li}}T - \left( Q_{il} + \frac{I_{il}}{d} \right) \frac{\delta S}{\delta Q_{lm}}T     \\ \nonumber
    &+& 2 \left( Q_{mi} + \frac{I_{mi}}{d}\right)\left( Q_{kl} + \frac{I_{kl}}{d}\right)\frac{\delta S}{\delta Q_{kl}}T  \bigg] \bigg( \lambda -1\bigg).
\end{eqnarray}

Next, the internal energy balance is expressed as

\begin{eqnarray}\label{evoleqeps}
\frac{\partial\epsilon}{\partial t} &=& -\frac{\partial}{\partial r_k} \left( v_k \, \epsilon\right) - \pi^S_{km} \, W_{km}  +  \frac{\delta E}{\delta Q_{kl}} R \left( \frac{1}{T} \frac{\delta E}{\delta Q_{kl}} - \frac{\delta S}{\delta Q_{kl}} \right)  
\\ \nonumber
&+& T\bigg(\lambda - 1 \bigg) \bigg[ 2 \left( Q_{ij} + \frac{I_{ij}}{d} \right) \frac{\delta S}{\delta Q_{ij}} \left( Q_{lk} + \frac{I_{lk}}{d} \right) W_{lk}  - \left( Q_{ik} + \frac{I_{ik}}{d} \right)  W_{kj} \, \frac{\delta S}{\delta Q_{ij}}  
\\ \nonumber
&-& \left( Q_{ik} + \frac{I_{ik}}{d} \right)  W_{jk} \, \frac{\delta S}{\delta Q_{ij}} \bigg] -  \frac{\delta E}{\delta \phi} \frac{1}{\rho} \frac{\partial}{\partial r_k} 
\rho D_{kl} \, \frac{\partial}{\partial r_l} \frac{1}{\rho} \,
\left( \frac{1}{T} \frac{\delta E}{\delta \phi} - \frac{\delta S}{\delta \phi} \right)  ,
\end{eqnarray}
the convection-diffusion, or a Generalized Cahn-Hilliard equation for the composition field is given by
\begin{equation}\label{evoleqphi}
\frac{\partial\phi}{\partial t} = - v_k \, \frac{\partial \phi}{\partial r_k} 
+  \frac{1}{\rho} \frac{\partial}{\partial r_k} 
\rho D_{kl} \, \frac{\partial}{\partial r_l} \frac{1}{\rho} \,
\left( \frac{1}{T} \frac{\delta E}{\delta \phi} - \frac{\delta S}{\delta \phi} \right)  ,
\end{equation}
and a convection-relaxation, or a Generalized BE equation for the orientation tensor prays
\begin{eqnarray}
\frac{\partial Q_{ij}}{\partial t} &=& - v_k \, \frac{\partial Q_{ij}}{\partial r_k} 
+ \left( \lambda \Xi_{ik} + \Omega_{ik}  \right) \left( Q_{kj} + \frac{I_{kj}}{d}  \right)  + \left( Q_{ik}  + \frac{I_{ik}}{d} \right) \left( \lambda \Xi_{kj} - \Omega_{kj}  \right)  
\nonumber \\ &-& 2\lambda \left(  Q_{ij}  + \frac{I_{ij}}{d} \right) \left(  Q_{kl} + \frac{I_{kl}}{d} \right) W_{lk} 
- R \left( \frac{1}{T} \frac{\delta E}{\delta Q_{ij}}
- \frac{\delta S}{\delta Q_{ij}} \right) ,
\label{evoleqQ}
\end{eqnarray}
with $W_{ij} = \partial v_i / \partial r_j$, $\Xi_{jk} = (W_{jk}+ W_{kj})/2$, and $\Omega_{jk} = (W_{jk}- W_{kj})/2$, representing the velocity gradient, strain rate, and vorticity tensors, respectively.

To conclude this section, we point out that the structure of the time-evolution equations remains unchanged as long as the number of $\bm{x}$ variables is the same. This holds even when new energetic and entropy contributions are added to Eqs.~(\ref{totalE}) and (\ref{totalS}), or existing ones are removed, as required by the physics of the system, for example in cases involving flexoelectric effects \cite{atzin2023minimal}, additional elastic and orientational terms, or, notably, different interface couplings. 

As an illustrative example where the number of $\bm{x}$ variables increases, we refer the reader to Appendix \ref{Multicomponent}, where we extend our model to the multi-component case. The generalization is straightforward: each component is labeled with a subscript $\Lambda$, to indicate its number, so that $\phi_\Lambda$ and $\bm{Q}_\Lambda$ represent the concentration and tensor order of that component, yielding $\bm{x} = (\rho^{mc}, \bm{m}^{mc}, \epsilon^{mc}, \phi_\Lambda, \bm{Q}_\Lambda)$, with the superscript $mc$ indicating a multi-component mixture variable.  There, we ultimately present the time-evolution equations of this system and discuss how to introduce a new constraint, namely that the sum of concentrations must always equal 1.

\section{Simulation Results}\label{Results}

We turn to numerical methods to examine the predictive capabilities of our GENERIC time-evolution equations against experimental observations. We use an upwind finite difference (FD) method combined with higher-order time integration. This approach follows the method of lines, where the differential equations (DEs) are first discretized in space and then integrated in time.

To implement these numerical techniques, we have developed the \texttt{LiquidCrystalGLB.jl} library in Julia (\url{https://github.com/depablogroup/LiquidCrystals.jl}). In this acronym, 'G' stands for GENERIC, and 'LB' for the Lattice Boltzmann Method for fluid dynamics. Our library efficiently handles derivative operators and other computational and mathematical tasks, while relying on the \texttt{DifferentialEquations.jl} package \cite{rackauckas2017differentialequations} for time integration. For additional implementation details, readers are referred to Appendix \ref{appendix:a}.

Unless otherwise stated, simulations are carried out in 2D Cartesian coordinates ($d = 2$), with the $x$- and $y$-axes defining the horizontal and vertical directions, respectively.

\subsection{Reaching Steady-State: Topological Defect Cores} \label{Equilibrium_integration}

Here, we present results for a lyotropic two-component system in the absence of flow. The terms in the Poisson operator $\bm{L}$ vanish due to the lack of convective mechanisms. Additionally, we neglect the energy balance, as it is uncoupled from the other DEs and is not a variable of interest at this stage. Thus, the equations of motion simplify to
\begin{equation}\label{CH_noflow}
\frac{\partial\phi}{\partial t} =  D_\rho \frac{\partial^2}{\partial {r_{l}}^2}   
\left( \frac{1}{T} \frac{\delta E}{\delta \phi} - \frac{\delta S}{\delta \phi} \right) ,
\end{equation}
and
\begin{equation}\label{GL_noflow}
\frac{\partial\bm{Q}}{\partial t} =- R \left( \frac{1}{T} \frac{\delta E}{\delta \bm{Q}}
- \frac{\delta S}{\delta \bm{Q}} \right) ,
\end{equation}
where we assume both the relaxation parameter $R$ and the parameter $D_\rho$ to be constant. The latter absorbs the (constant) total density dependence and has units of [length$^5$/(entropy$\cdot$time)].

To find the steady-state values of Eqs. (\ref{CH_noflow}) and (\ref{GL_noflow}), which ensure dynamic stability, we set the time derivatives to zero ($\partial \bm{Q} / \partial t = \bm{0}$ and $\partial \phi / \partial t = 0$) \cite{matsuyama2002non}. This results in the following equilibrium conditions:

\begin{equation}\label{phieq}
   \frac{1}{T} \frac{\delta E}{\delta \phi} - \frac{\delta S}{\delta \phi}  = 0 ,
\end{equation}
and 
\begin{equation}\label{q_eq}
   \frac{1}{T} \frac{\delta E}{\delta q} - \frac{\delta S}{\delta q}  = 0.
\end{equation}
Note that Eq. (\ref{q_eq}) is written in terms of $q$, rather than $\bm{Q}$, with $Q_{ij}Q_{ji}=q^2/2$ and $Q_{ij}Q_{jk}Q_{ki}=0$ in 2D, given the expected correlation between the $\phi$ and $q$. Specifically, we anticipate that when the concentration exceeds the critical threshold for phase transition, a 'high' value, the order parameter should enter the nematic range, typically  $q\geq 0.6$. More explicitly, these conditions, along with Eqs.~(\ref{totalE}) and (\ref{totalS}), become:
\begin{equation}  \label{phi_equilbrium}
      \chi(1-2\phi)+T\left[C_1\ln{\phi}-C_2\ln{(1-\phi)}+C_1-C_2\right] -2A_0U\phi\left(\frac{1}{12}q^2-\frac{1}{16}q^4\right)+ \frac{1}{4}A_0Tq^2 = 0,
\end{equation}
and
\begin{equation}  \label{S_equilbrium}
     -\frac{1}{2}\rho A_0\phi q^2\left[\frac{\phi U}{3}-\frac{\phi Uq^2}{2}-T\right] = 0,
\end{equation}
where we assume a homogeneous steady-state, implying that spatial derivatives vanish in the bulk ($\partial \bm{Q} / \partial \bm{r} = \bm{0}$ and $\partial \phi / \partial \bm{r} = 0$ as $\bm{r} \to \infty$), corresponding to the mean-field limit. 

Moreover, the equilibrium conditions adopt these forms for the LdG functions, consistent with the discussion in the previous section and common literature notation \cite{Kumareaat7779, Sulaimanetal06, zhang2021spatiotemporal}:
\begin{equation}  \label{A_phi}
     A(\phi) = A_0 \phi,
\end{equation}
and
\begin{equation}  \label{B_phi}
     B(\phi) = A_0 U \phi^2,
\end{equation}
where $A_0$ is a constant parameter controlling the strength of both the entropic and energetic LdG contributions, and $U$ is a nematic potential related to the anisotropic particle density \cite{doi1981molecular}.

\begin{figure}[hbt!]
	
		\centering
		\includegraphics[scale=0.45]{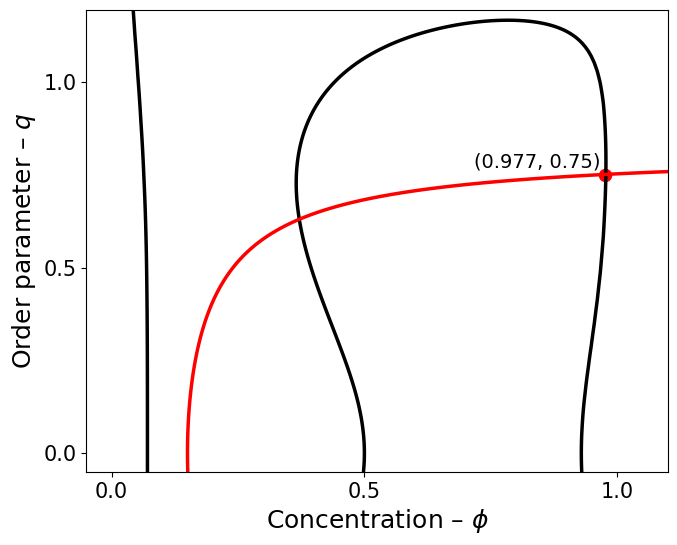}

	\caption{Plots of Eq. \ref{phi_equilbrium} (in black) and Eq. \ref{S_equilbrium} (in red) using parameters $T=1$, $C_1 = C_2 =1$, $\chi= 3$, $A_0=1$, and $U=20$. $x$-axis denotes the concentration range, while $y$-axis indicates the order parameter. One of the intersection points is indicated with a red dot, denoting $\phi_e$ and $q_e$}
 \label{Eq_Curves}
	
\end{figure}

Eqs. (\ref{phi_equilbrium}) and (\ref{S_equilbrium}) are plotted in Fig. \ref{Eq_Curves}. To manage the extensive degrees of freedom, we set $\rho=1$, $T=1$, $C_1 = C_2 =1$, $\chi= 3$, and $A_0=1$, which are standard \cite{kruger2017lattice, head2024spontaneous, CHEN2019100031, CHAN1995377}. We also choose $U=20$ to ensure that both curves intersect near $\phi_e = 0.977$ and $q_e=0.75$, which serve as our equilibrium or threshold values for concentration and order, respectively. For a mapping of these, and subsequent, computational values to physical quantities, please refer to Appendix \ref{appendix:c}. 

\begin{figure}[hbt!]
    \centering
    
    \begin{subfigure}{0.8\textwidth}
        \centering
        \includegraphics[scale=0.6]{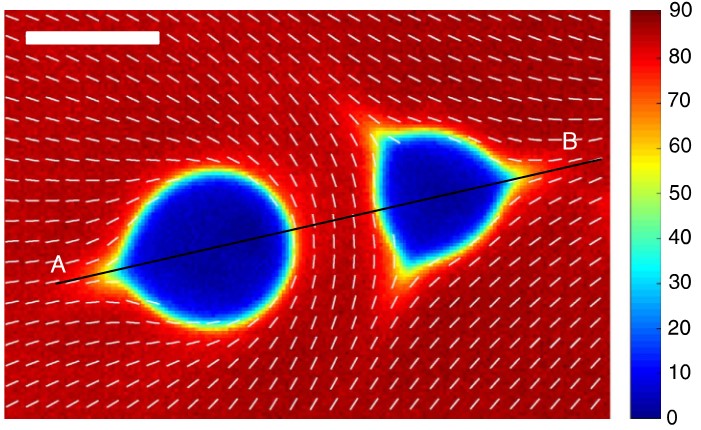}
        \caption{}
        \label{Lav_exp}
    \end{subfigure}
    
    \vspace{1em} %
    
    \begin{subfigure}{0.8\textwidth}
        \centering
        \includegraphics[scale=0.22]{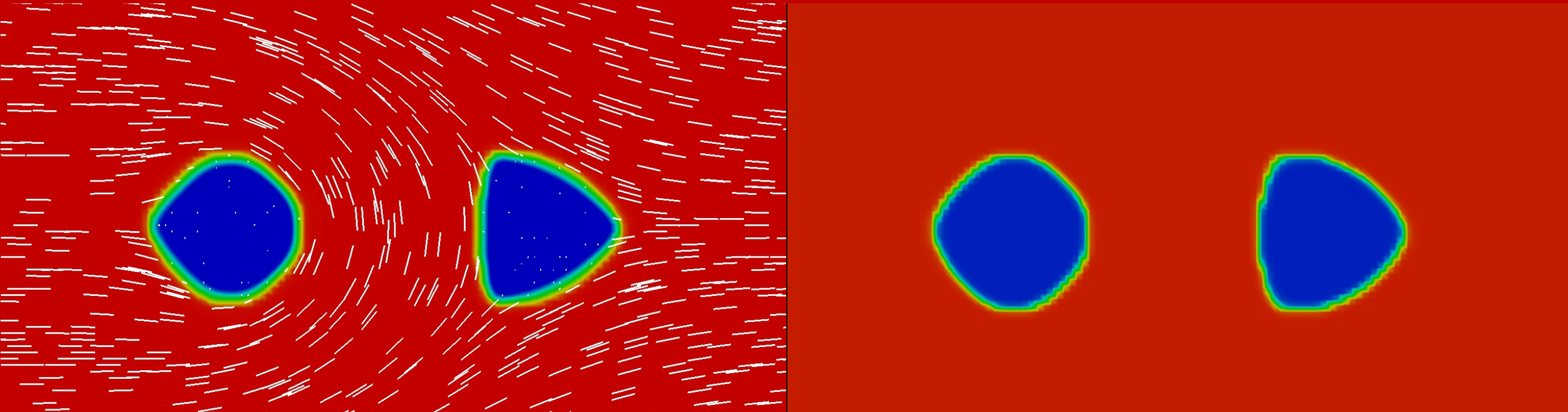}
        \caption{}
        \label{Lav_Final}
    \end{subfigure}

    \begin{subfigure}{0.45\textwidth}
    \centering
    \includegraphics[scale=0.42]{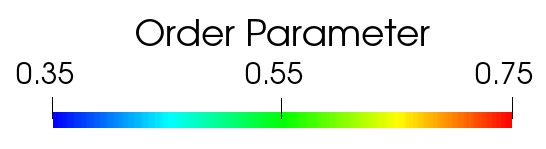}
    \caption{}\label{Lav_Order_colorbar}
\end{subfigure}
\hfill
\begin{subfigure}{0.45\textwidth}
    \centering
    \includegraphics[scale=0.42]{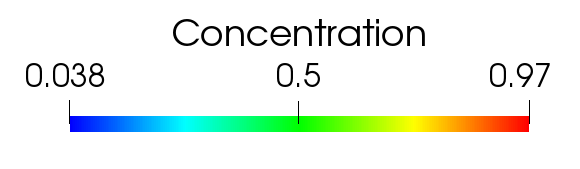}
    \caption{}\label{Lav_conc_colorbar}
\end{subfigure}
    
    \caption{Comparison of topological core defects in a nematic medium. (\subref{Lav_exp}) Experimental birefringence map of a DSCG mixture in a thin cell with tangential anchoring (scale bar: 20~$\mu$m), showing optical retardance (color bar) and director orientation (white lines); adapted from \cite{zhou2017fine}. (\subref{Lav_Final}) Final-time-step simulation depicting the director field (white lines, scaled according to the concentration profile), the scalar order parameter (left), and the concentration distribution (right), with their corresponding color bars shown in (\subref{Lav_Order_colorbar}) and (\subref{Lav_conc_colorbar}).}
    \label{fig:Lav_exp_sim}
\end{figure}

Next, we present the experimental data that inspire our first numerical result. Zhou et al. studied a disodium cromoglycate (DSCG) liquid crystal–water mixture at 15\% by weight, which exhibits a nematic phase at room temperature \cite{zhou2017fine}. When confined between glass plates imposing degenerate tangential anchoring, this mixture shows topological core patterns. Two $\pm1/2$ tactoid defects appear in the DSCG-depleted regions, each displaying sharp peaks: one for the +1/2 defect and three for the -1/2 (Fig. \ref{Lav_exp}). To simulate this,  we use a structural mesh with uniform spacing in both directions, $\Delta x=\Delta y=1.0$, and periodic boundary conditions (PBCs).

The simulation domain has size $(N_x , N_y) = (200, 200)$, where $N_x$ and $N_y$ denote the number of nodes in the $x-$ and $y-$ directions, respectively. The initial conditions (ICs) are chosen near equilibrium, with a director field resembling the experimental pattern, such that the orientation of $\bm{n}$ triggers the emergence of topological cores. The field $q$ is homogeneously initialized at $q_e=0.75$. Meanwhile, the initial concentration profile consists of two rounded droplets of radius $r_d = 15 \Delta x$, centered at $\left(N_x/2 \pm (r_d + 18\Delta x\right)  , N_y/2)$. Each droplet has a concentration of $\phi = 0.05$, associated with a low order parameter, and is immersed in a high-concentration environment near equilibrium ($\phi \approx 0.97$). In other words, these are two isotropic droplets surrounded by a nematic phase (e.g., water within DSCG). For a snapshot of the ICs, see the Supplemental Material \cite{SI} (Video S1, first frame).

In selecting parameters for spatial gradient terms, we follow the guideline that $\kappa_E \approx \Delta x$ to ensure numerical stability \cite{DEMELLO2005429}, and apply this principle to other parameters as well. After tuning, we settle on $\kappa_E = 1.0 $, $K = 1.5$ (fixed constant for practicality) and $ \kappa'= 1.5$, which, as a reminder, also enforces tangential (planar) anchoring near the interface. For simplicity, we neglect the entropic effect of diffuse interfaces by setting $\kappa_S = 0$. Furthermore, we set $R= 0.1$ and $D_\rho = 0.0075$, consistent with characteristic values for practical time integration, with $D_\rho$ specifically adjusted to keep $\phi$ within its physical range. We complete our setup by selecting the timestep $\Delta t = 1.0$ for the \texttt{CarpenterKennedy2N54} time integrator (from \texttt{DifferentialEquations.jl}), and evolving the system to its final number of time steps $t_f =120,000$.

As time progresses, Video S1 \cite{SI} shows the evolution of the three fields ($\bm{n}$, $q$, and $\phi$), and Fig. \ref{Lav_Final} depicts the final step. The $\bm{n}$ field becomes smoother and stabilizes the topological defect cores. Notably, the initially circular $\phi$ profile evolves to match $q$, deforming the droplets into tactoid-like structures. This deformation arises naturally from the chosen parameters; otherwise, the droplets tend to vanish. An interfacial region also forms around the tactoids, consistent with experimental observations. Outside the defect cores, both scalar profiles remain close to their equilibrium values ($\phi_e$ and $q_e$).

Our simulations produce stable simulations that preserve physical constraints, maintaining $\phi$ and $q$ within $(0, 1)$ and a unitary director field. This approach successfully reproduces the experimental pattern. 

Note that for the calculations presented here we assume a constant $K$. However, we hypothesize that using a $\phi$-dependent $K$—as suggested for more realistic scenarios \cite{PerezLemus2017}—particularly in the form of a sigmoidal function mimicking the diffuse interface, could lead to droplet swelling. Based on the model structure, we expect both $\kappa_E$ and constant $K$ to govern the thickness of the isotropic-nematic interface, as both are associated with squared gradients. We also hypothesize that $\kappa'$ modulates the sharpness of the topological peaks and the speed at which $\phi$ and $q$ become correlated. These hypotheses require further analytical and numerical investigation.

\subsection{LC Droplets under Controlled Flow} \label{Convection-Diffusion-Relaxation}

Next, we introduce hydrodynamics into our equations via a controlled flow, meaning that the velocity profile is externally imposed, and again compare our simulation results with experimental data. 
In this scenario, Eqs. (\ref{evoleqphi}) and (\ref{evoleqQ})  can be expressed as:
\begin{equation}\label{conv_diff_eq}
\frac{\partial\phi}{\partial t} + v_k  \frac{\partial \phi}{\partial r_k}  = D_\rho \frac{\partial^2}{\partial {r_{l}}^2}
\left( \frac{1}{T} \frac{\delta E}{\delta \phi} - \frac{\delta S}{\delta \phi} \right),
\end{equation}
and
\begin{equation} \label{conv_relax_eq}
\frac{\partial\bm{Q}}{\partial t} + \bm{v} \cdot \frac{\partial \bm{Q}}{\partial \bm{r}}  = -  R \left( \frac{1}{T} \frac{\delta E}{\delta \bm{Q}}
- \frac{\delta S}{\delta \bm{Q}} \right),
\end{equation}
which represent a Convection-Diffusion-Relaxation system of DEs. Note that we neglect the velocity gradients ($\bm{W}$) of equation (\ref{evoleqQ}) since these gradients mainly affect tensorial variables such as $\bm{Q}$ \cite{macosko1994rheology}; in contrast, $q$, a scalar and the main variable of interest here, does not require them for transport. Consequently, we allow $\bm{n}$ to be rearranged during the relaxation process. Moreover, we consider only two velocity profiles: one with constant velocities, where $\bm{W}$ vanishes everywhere, and another with a parabolic profile. In the latter, flow varies along the $y$-direction, but our focus lies near the center of a wide parabola, where the velocity is maximal and $\bm{W}_{y= N_y /2} = \bm{0}$. Elsewhere, contributions from $\bm{W}$ remain insignificant due to the smooth and gradually varying profile. 

The prescribed velocity fields serve as controlled test cases, chosen for their well-known forms to avoid solving the full momentum balance. We assume that the conditions required to generate these flows are imposed externally (e.g., via pressure gradients or wall motion) and are \emph{fully developed} at the onset of the simulation for modeling convenience. As before, we use PBCs (e.g., we model a channel with translational symmetry).

In the absence of diffusion and relaxation ($D_\rho=0$ and $R=0$, or with $\bm{M}^{\rm diff}_{\phi}$ and $\bm{M}^{\rm relax}_{\bm{Q}}$ omitted), Eqs. (\ref{conv_diff_eq}) and (\ref{conv_relax_eq}) reduce to pure advection equations, for which many elaborate methods exist, including higher-order spatial differentiation and averaging \cite{leveque2007finite}. Here, we employ a simpler splitting approach: the advection terms are handled separately using a first-order (accurate) upwind scheme and then incorporated into the diffusion-relaxation solver described in Section \ref{Equilibrium_integration}, which employs a second-order central FD approximation (see Appendix \ref{appendix:a}).

The combined sequence, advection followed by diffusion-relaxation, constitutes one overall time step, repeated until reaching the final simulation time or the specified number of advective steps. The basic elements for the time integration are provided in Fig. \ref{Flowchart} of Appendix \ref{Appex:Flowchart}. This method introduces two distinct computational time scales: a short $\Delta t_{adv} $ for advection and a longer $\Delta t_{dr}$ for diffusion-relaxation. Although diffusion-relaxation is considered to be physically slower, it requires numerical acceleration under flow conditions to keep pace with the convective time scale. Typically, we maintain $\Delta t_{dr} \geq 2\Delta t_{adv} $, but empirically adjust this ratio based on expected physics. For example, we speculate that it may relate to specific rheological properties, such as breaking point length or droplet elasticity. 

\sloppy

Next, we discuss the 2D Courant–Friedrichs–Lewy (CFL) condition in upwind schemes \cite{balsara2013, leveque2007finite}
\begin{equation} \label{CFL_condition}
    C_{FL} = \Delta t_{adv} \left( \frac{\hat{v}_x}{\Delta x} + \frac{\hat{v}_y}{\Delta y}  \right) \leq C_{max},
\end{equation}
where $C_{FL}$ denotes the dimensionless CFL number, $\hat{v}$ represents the absolute value of the velocity component in the direction indicated by its subscript, and $C_{max}$ is a threshold used to ensure numerical stability. For an explicit time-marching solver, such as the Euler method used in our upwind scheme, $C_{max} = 1$. In particular, since our flows are restricted to a single direction, the CFL condition simplifies to 
\begin{equation} \label{CFL_1D}
    \hat{v}_x \leq \frac{\Delta x}{\Delta t_{adv}} .
\end{equation}
Following this condition prevents spurious or oscillatory instabilities, and we use it to set the numerical parameters accordingly. Nonetheless, some numerical error is inevitable, as the scalar variables may reach unphysically high values with each iteration \cite{balsara2013}. To counteract these effects, we rely on the diffusion-relaxation part to restore $q$ and $\phi$ to their physical (equilibrium) values. This also updates the director vectors to their new configuration, as dictated by the velocity profile, implying an intermediate Ericksen number \cite{fedorowicz2023effects}.   

\begin{figure*}[hbt!]
    \centering
        \begin{subfigure}[t]{0.49\textwidth}
            \includegraphics[scale=0.228]{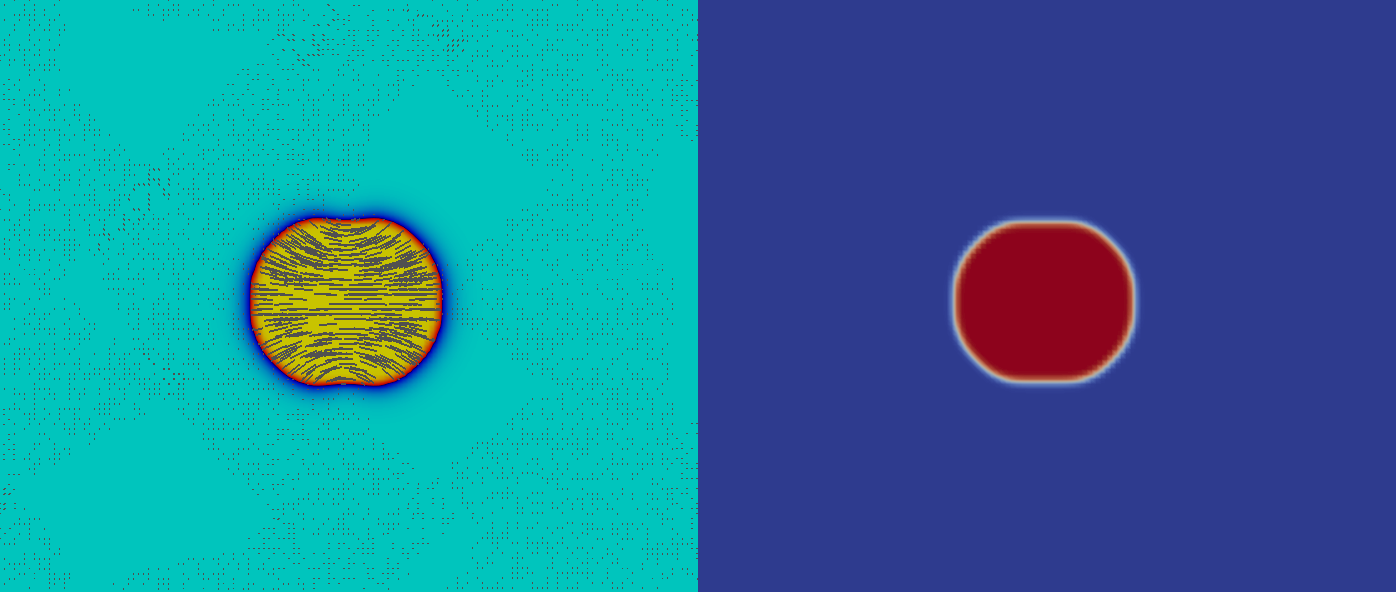}
            \caption{}\label{Couette_IC}
            
        \end{subfigure}%
        \hfill
        \begin{subfigure}[t]{0.49\textwidth}
            \includegraphics[scale=0.135]{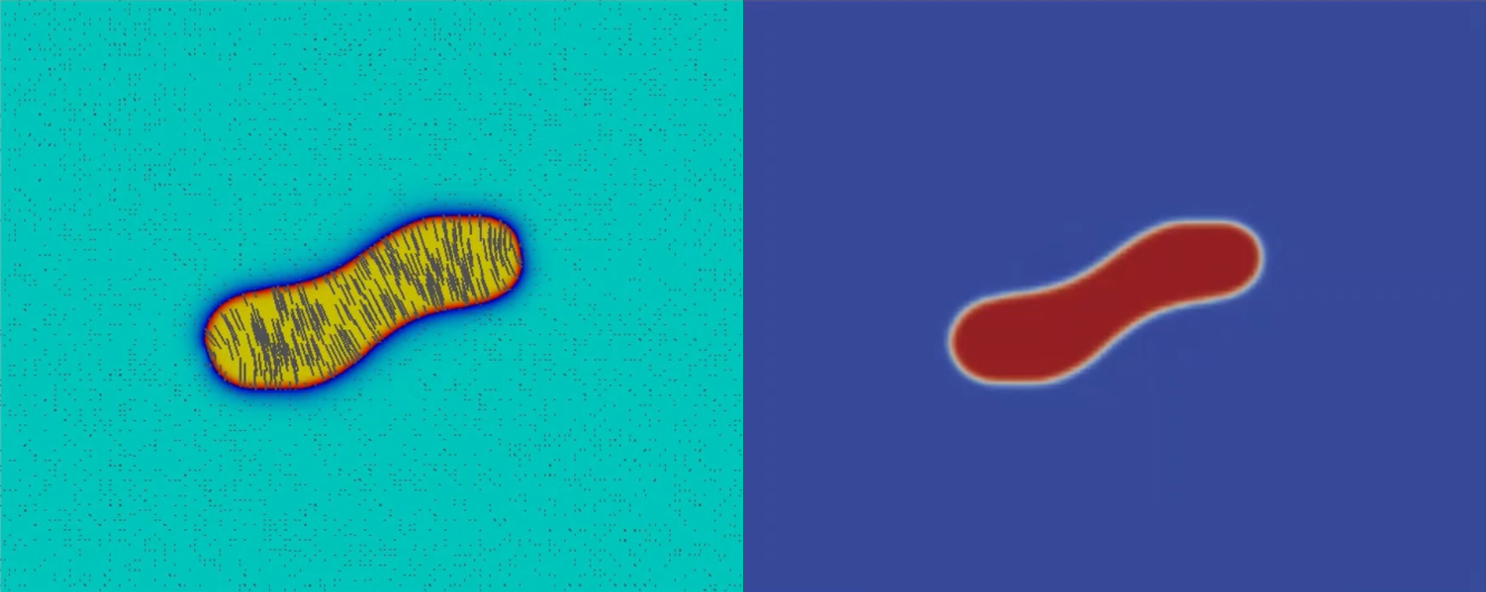}
            \caption{}\label{Couette_Snap1}
        \end{subfigure}

        \vspace{0.5em}

        \begin{subfigure}[t]{0.49\textwidth}
            \includegraphics[scale=0.135]{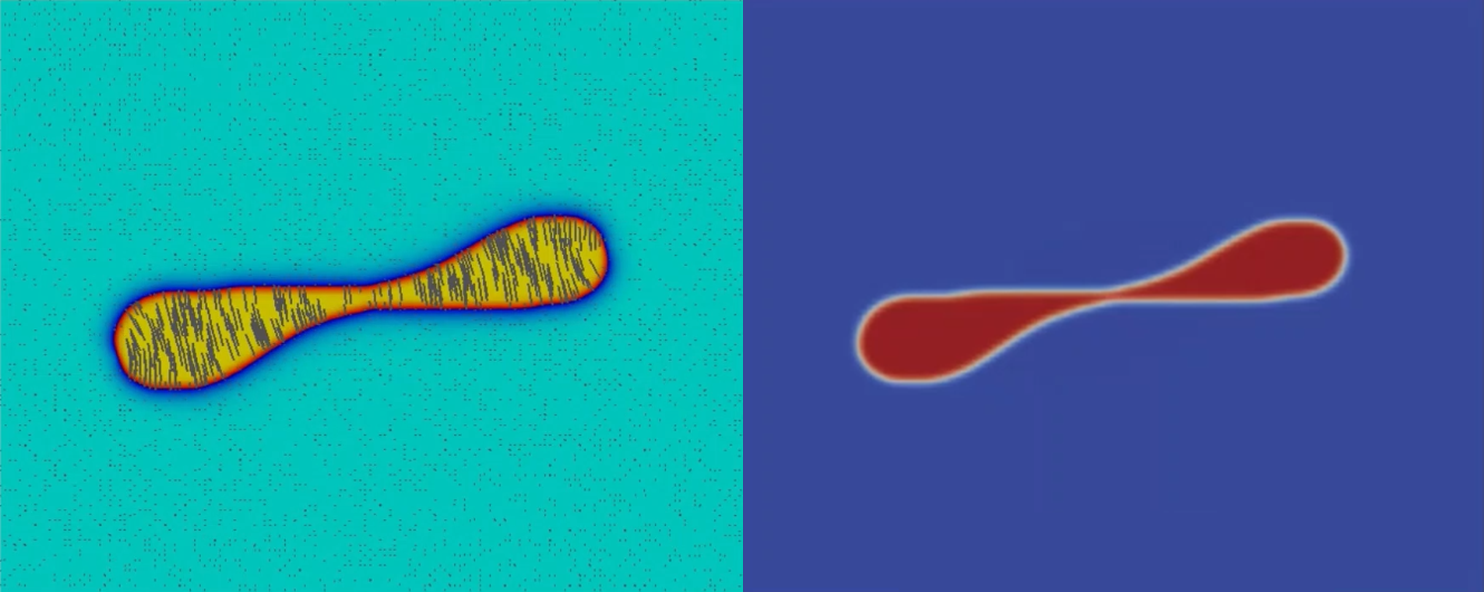}
            \caption{}\label{Couette_Snap2}
        \end{subfigure}%
        \hfill
        \begin{subfigure}[t]{0.49\textwidth}
            \includegraphics[scale=0.135]{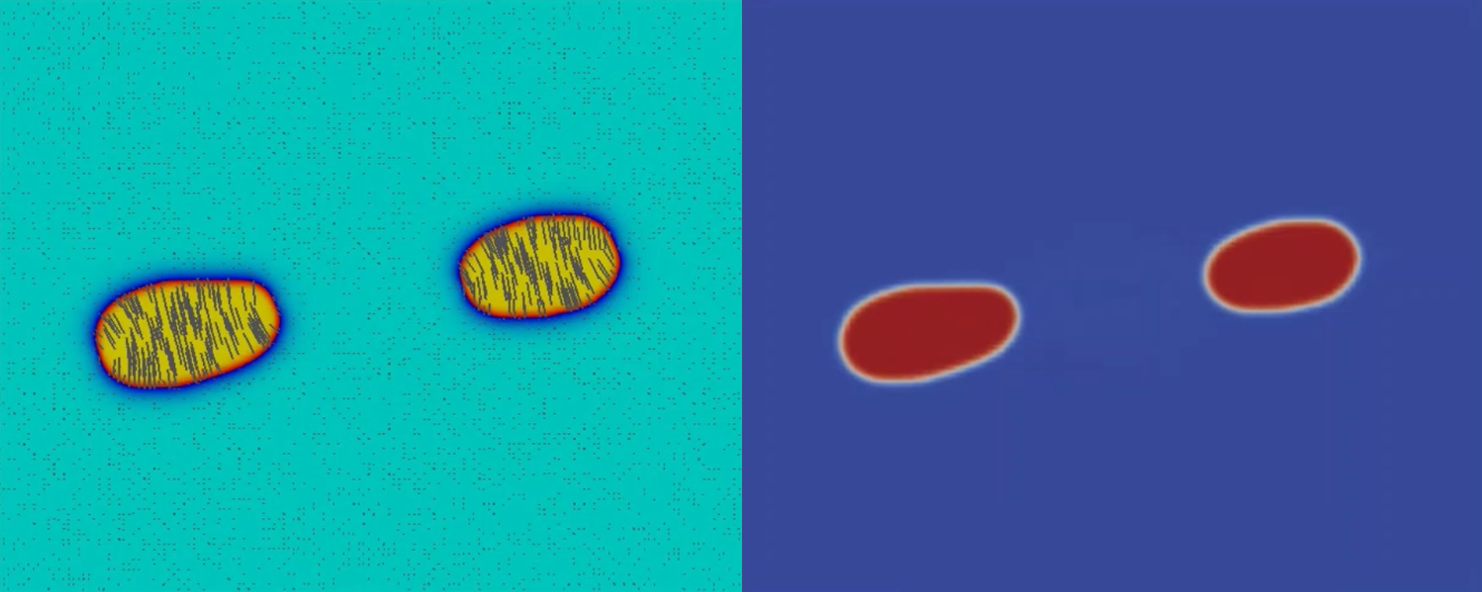}
            \caption{}\label{Couette_Final}
        \end{subfigure}

        \vspace{0.5em}

            \begin{subfigure}[t]{0.48\textwidth}
                \includegraphics[scale=0.45]{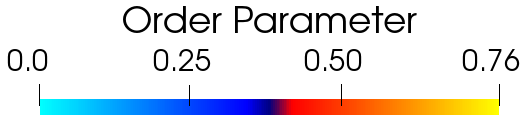}
                \caption{}\label{Couette_colorbar_order}
            \end{subfigure}%
            \hfill
            \begin{subfigure}[t]{0.48\textwidth}
                \includegraphics[scale=0.45]{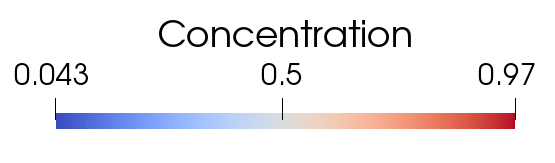}
                \caption{}\label{Couette_colorbar_concentration}
            \end{subfigure}

    \vspace{0.5em}
    \begin{subfigure}[t]{0.48\textwidth}
        \centering
        \hspace{3.5em} 
        \includegraphics[scale=0.13]{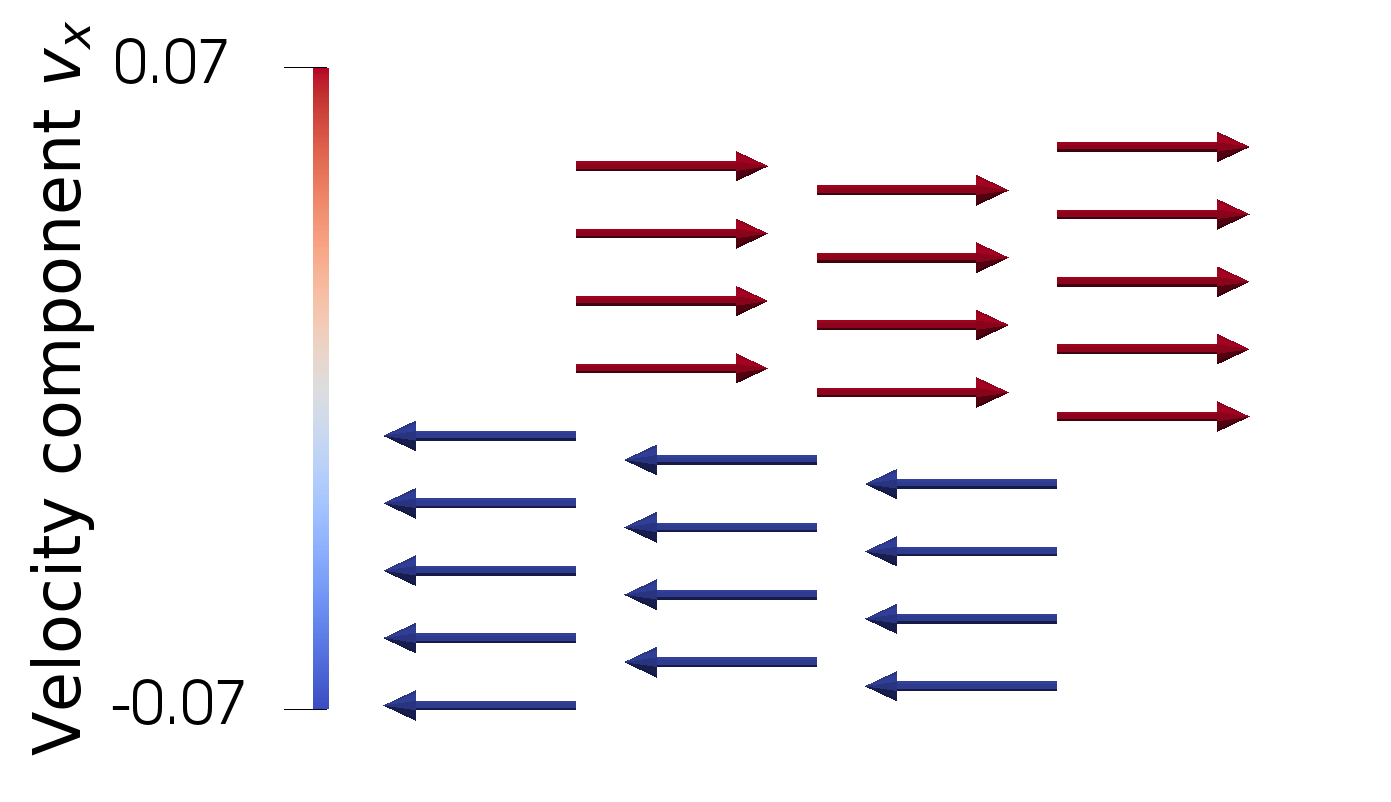}
        \caption{}\label{Couette_velocity_profile}
    \end{subfigure}
    \hfill
    \begin{subfigure}[t]{0.48\textwidth}
    \centering
    \includegraphics[scale=0.95]{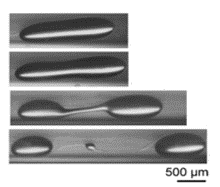}
    \caption{}\label{Couette_exp}
    \end{subfigure}

    \caption{Comparison of an axial droplet under semi-Couette flow. Subfigures (\subref{Couette_IC})--(\subref{Couette_Final}) show simulation snapshots: in each panel, the left side displays the director field (gray lines, scaled by the order parameter) and the scalar order parameter, while the right side depicts the concentration field. (\subref{Couette_IC}) displays the initial conditions. Subfigures (\subref{Couette_colorbar_order}) and (\subref{Couette_colorbar_concentration}) provide the corresponding color bars, while (\subref{Couette_velocity_profile}) presents the schematic representation of the velocity profile imposed in the simulation. (\subref{Couette_exp}) compares the experimental evolution of a poly(isobutylene) (PIB, Parabol 1300) droplet deformation under shear flow; adapted from \cite{Vananroye2006}.}
    \label{Couette_sim_exp}
\end{figure*}

As an illustrative case, we study the evolution of an LC droplet with axially aligned directors under semi-Couette flow. We use this name because, unlike the classical Couette flow with linear $y$-dependence, our velocity profile maintains a constant magnitude but is oppositely directed in the upper and lower halves of the domain (schematic representation in Fig. \ref{Couette_velocity_profile}). This profile remains 'active' until droplet breakup, after which it decays to zero. Using previous mesh parameters and setting $\Delta t_{adv} = 15$, Eq. (\ref{CFL_1D}) yields $\hat{v}_x = 0.0663$ for both regions of the simulation box. 

Fig. \ref{Couette_IC} displays the ICs for $\phi$ and $\bm{Q}$, an axial droplet \cite{cryst14110934}, precomputed with the diffusion-relaxation solver, demonstrating the robustness of our framework in reproducing a variety of LC configurations. This droplet is centered in the computational domain ($r_d=13 \Delta x$, $\phi = \phi_e$, $q = q_e$) and immersed in an isotropic medium ($q\approx0$, $\phi= 0.05$). Here, we take perpendicular anchoring ($\kappa' = -0.85$), a defining characteristic of axial droplets, indicated by its negative value. All other computational parameters remain the same as in section \ref{Equilibrium_integration}, except those related to time, detailed next. 

We evolve the system over a total number of 45 advective steps. In each advection step, we apply $\Delta t_{dr}=0.8\times 500=400$, based on the solver parameters $\Delta t=0.8$ and $t_f=500$ used in the \texttt{CarpenterKennedy2N54} time integrator.

As time progresses, Video S2 \cite{SI} shows the evolution of the three fields ($\bm{n}$, $q$, and $\phi$), and  Fig.~\ref{Couette_sim_exp}\subref{Couette_Snap1}-\subref{Couette_Final} depicts selected snapshots. The droplet begins to elongate, and the motion continues until a small neck forms and eventually breaks, producing two droplets. We can now also track the evolution of the director vectors, which realign due to the droplet’s new configuration. To our knowledge, no analogous LC experiment exists. However, for completeness, Fig.~\ref{Couette_exp} shows an emulsion of poly(isobutylene) (PIB, Parabol 1300)—not liquid crystalline—under Couette flow that resembles our simulation results.

As an additional example, we examine the evolution of the same axial droplet under Poiseuille (parabolic) flow. Again, most considerations listed above are unchanged. The main difference lies in the time steps and the velocity field used here.

\begin{figure}[H]
    \centering
    
    \begin{subfigure}{0.48\textwidth}
        \centering
        \includegraphics[scale=0.26]{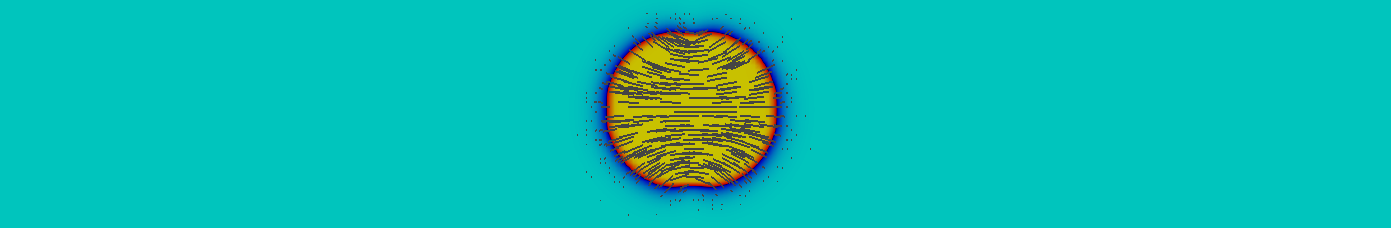}
        \caption{}\label{Parabolic_IC_Q}
        \includegraphics[scale=0.26]{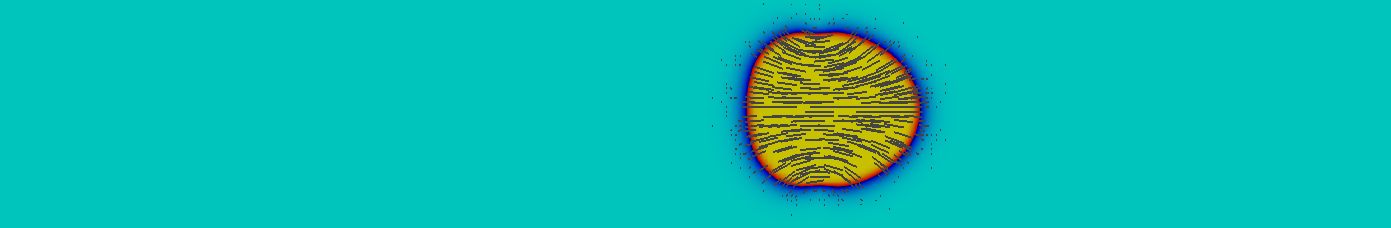}
        \caption{}\label{Parabolic_Snap1_Q}
        \includegraphics[scale=0.26]{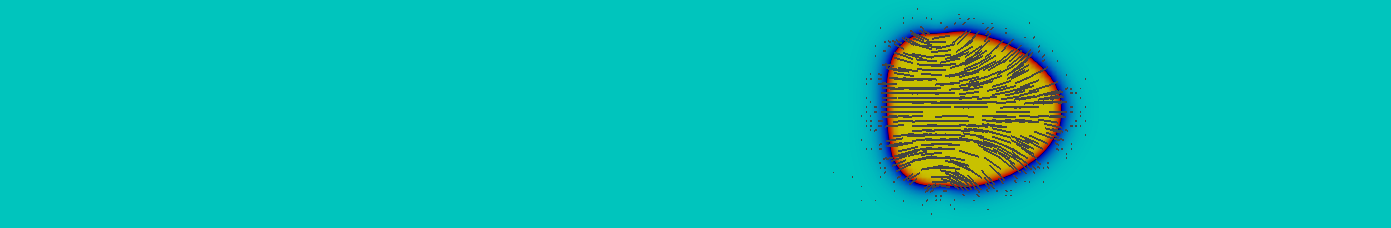}
        \caption{}\label{Parabolic_Snap2_Q}
        \includegraphics[scale=0.26]{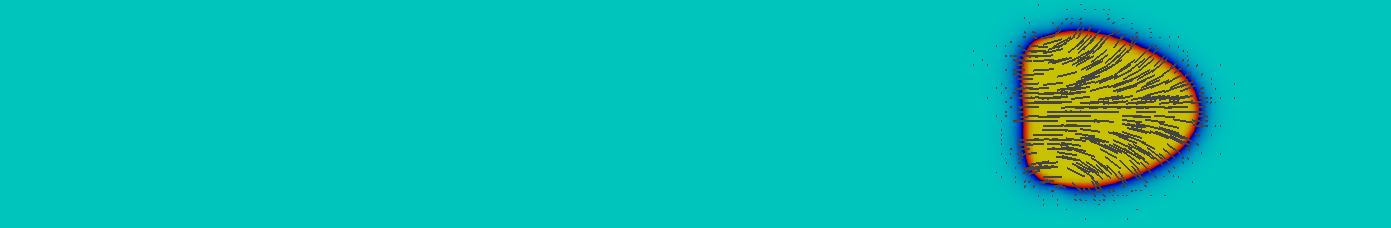}
        \caption{}\label{Parabolic_Final_Q}

    \end{subfigure}
    \hfill
    \begin{subfigure}{0.48\textwidth}
        \centering
        \includegraphics[scale=0.26]{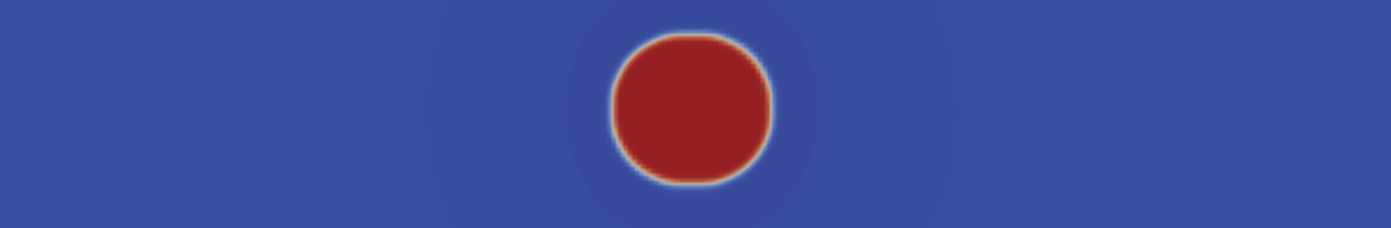}
        \caption{}\label{Parabolic_IC_phi}
        \includegraphics[scale=0.26]{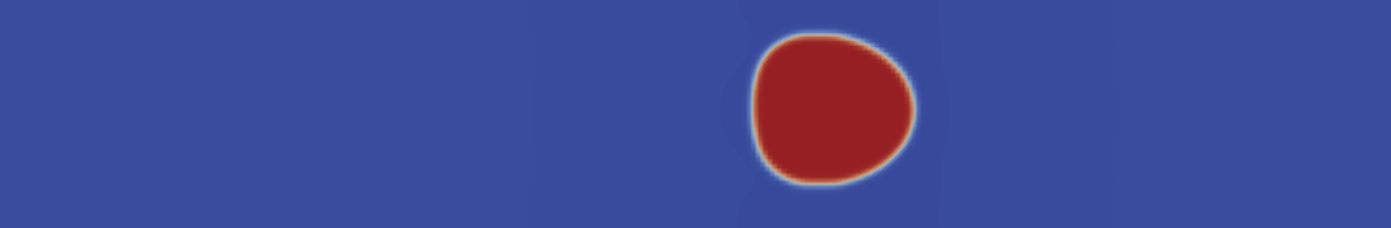}
        \caption{}\label{Parabolic_Snap1_phi}
        \includegraphics[scale=0.26]{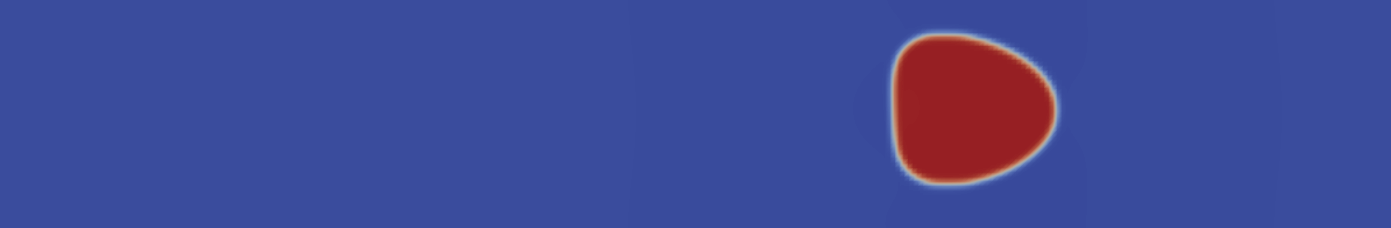}
        \caption{}\label{Parabolic_Snap2_phi}
        \includegraphics[scale=0.26]{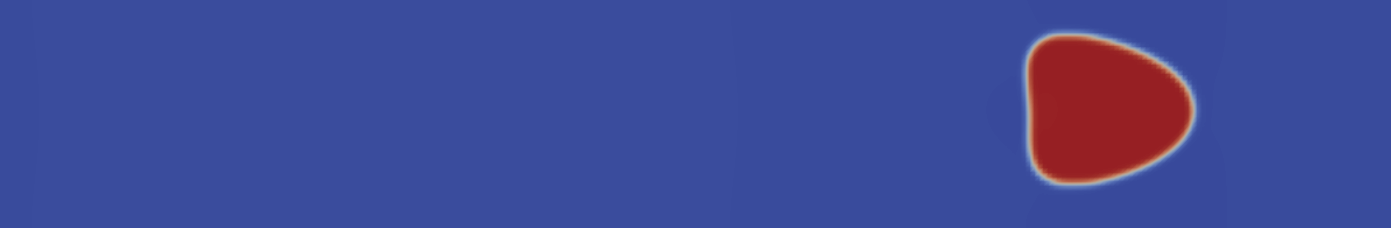}
        \caption{}\label{Parabolic_Final_phi}
  
    \end{subfigure}

    \vspace{1em}
    \begin{subfigure}{0.48\textwidth}
        \centering
        \includegraphics[scale=0.45]{images/Colorbar_orderparameter.png}
        \caption{}\label{parabolic_color_order}
       
    \end{subfigure}
    \hfill
    \begin{subfigure}{0.48\textwidth}
        \centering
        \includegraphics[scale=0.45]{images/Colorbar_concentration2.png}
        \caption{}\label{parabolic_color_phi}
       
    \end{subfigure}
    
    \vspace{1em}
    \begin{subfigure}{0.48\textwidth}
        \centering
        \includegraphics[scale=0.21]{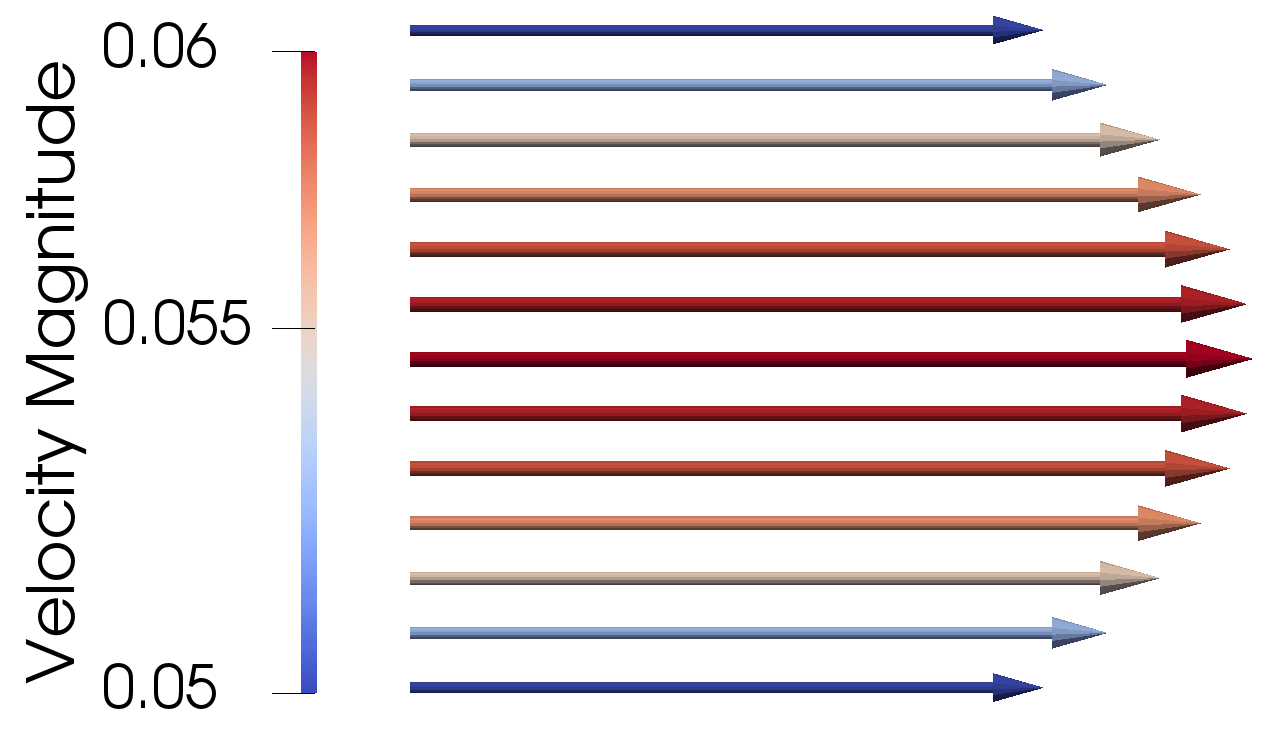}
        \caption{}\label{Parabolic_velocity_profile}
    \end{subfigure}
    \hfill
    \begin{subfigure}{0.48\textwidth}
    \centering
    \includegraphics[scale=0.19]{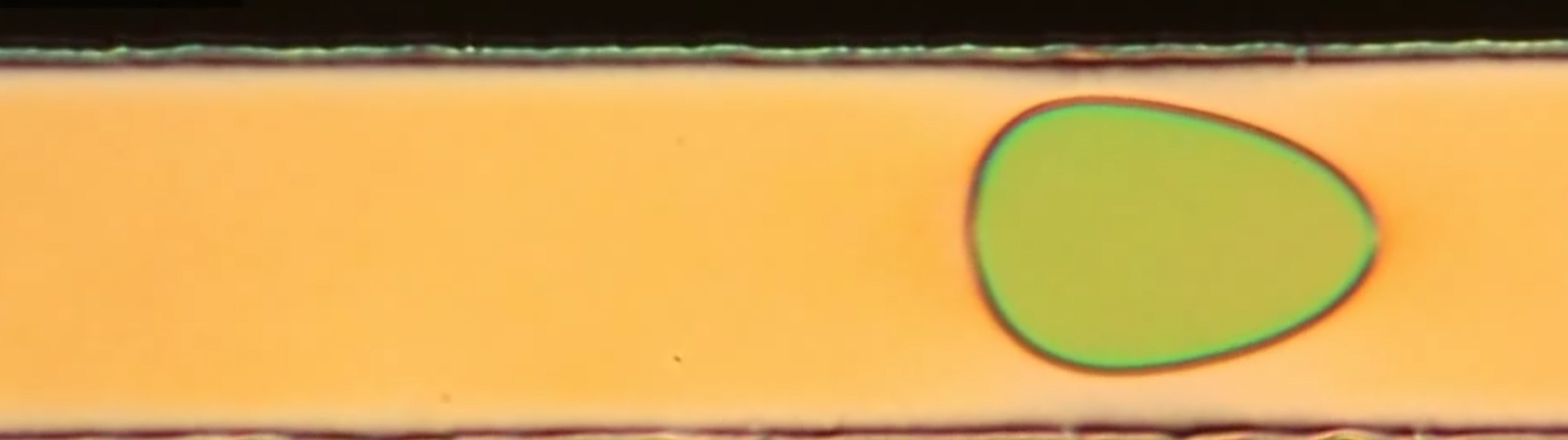}
    \caption{}\label{Parabolic_exp}
    \end{subfigure}

    \caption{Comparison of an axial droplet under parabolic flow. 
Subfigures (\subref{Parabolic_IC_Q})--(\subref{Parabolic_Final_phi}) present simulation snapshots: (\subref{Parabolic_IC_Q})--(\subref{Parabolic_Final_Q}) depict the director field (gray lines, scaled by the order parameter) and the scalar order parameter, while (\subref{Parabolic_IC_phi})--(\subref{Parabolic_Final_phi}) show the corresponding concentration fields. Initial conditions are in (\subref{Parabolic_IC_Q}) and (\subref{Parabolic_IC_phi}), intermediate time steps in (\subref{Parabolic_Snap1_Q}), (\subref{Parabolic_Snap2_Q}), (\subref{Parabolic_Snap1_phi}), and (\subref{Parabolic_Snap2_phi}), and the final state in (\subref{Parabolic_Final_Q}) and (\subref{Parabolic_Final_phi}). Subfigures (\subref{parabolic_color_order}) and (\subref{parabolic_color_phi}) display the respective color bars. (\subref{Parabolic_velocity_profile}) shows the parabolic velocity profile imposed in the simulation. (\subref{Parabolic_exp}) presents an experimental still of a 5CB droplet ($\sim$20~$\mu$m) deformed in a pressure-driven microfluidic flow and observed under a polarized microscope; adapted from \cite{Tadej2019Sculpting}.
}
    \label{Parabolic_exp_sim}
\end{figure}

To resemble a local segment of the classical planar Poiseuille flow, we choose a profile described by the shifted analytical form
$ v_x(y) =  V_{\text{max}} \left( \frac{y}{H}  - \frac{y^2}{H^2}  + \frac{3}{4} \right) $,
where \( H=50 \) is a design parameter used to flatten the parabola, associated with the channel height (here $N_y$). The added $3/4$ term ensures that $ V_{\text{max}} $, the maximum velocity, occurs at $y=N_y/2$. This results in a plug-like flow that still preserves the underlying parabolic profile (see Fig. \ref{Parabolic_velocity_profile}). With $\Delta t_{adv} = 15$ again, we set $ V_{\text{max}} = 0.062$ to satisfy the CFL condition. Additionally, the box size is now taken as $(N_x , N_y) = (300, 50)$ to give the appearance of a channel. We then run the simulation for 99 advective time steps and use $\Delta t_{dr}=32$ to prevent the droplet from excessively elongating.

Fig. \ref{Parabolic_exp_sim}\subref{Parabolic_IC_Q}-\subref{Parabolic_Final_phi} displays the evolution of the three fields across selected frames; the full time course is available in Video S3 \cite{SI}. We observe that the droplet gradually deforms into a triangular shape, given by the velocity profile. The director vector realigns and curves around the vertices of this triangle while pointing perpendicular to the edges. As far as we are aware, no exact LC experiment reproduces our simulation; however, Fig. \ref{Parabolic_exp} presents a nematic LC droplet of 4-Cyano-4'-pentylbiphenyl (5CB), induced by temperature gradients (thermotropic rather than lyotropic), surrounded by another nematic phase and subjected to a similar parabolic flow. Notably, the experimental nematic droplet again mirrors the dynamics of our simulation.

Taken together, our results indicate that our proposed formalism is capable of reproducing experimental observations for lyotropic systems far from equilibrium.

\section{Conclusions}\label{Conclusions}

This work presents a comprehensive theoretical framework for lyotropic LCs built upon the GENERIC formalism, which naturally ensures thermodynamic consistency out of equilibrium. Starting from a binary mixture, we derived time-evolution equations for mass, momentum, energy, concentration, and liquid crystalline order, yielding additional momentum contributions and incorporating thermal effects and energy balances that are not commonly addressed in the literature. We then extended this formalism to obtain a multi-component generalization.

Our Julia-based solver, employing custom numerical schemes, enables stable and efficient simulations of binary LC mixtures, from equilibrium topological defects to flow-driven droplet shape transitions, achieving good agreement with experimental data. The computational framework accounts for a set of constraints, including concentration limits, the CFL condition, and multiple time scales. In a follow-up study, we plan to solve the full set of four coupled equations to demonstrate the emergence of active turbulent lyotropic LCs. 

Overall, this work bridges theory, computation, and experiment, offering a predictive platform not only for LC systems but also for more complex multicomponent mixtures, with potential applications in materials science, bioengineering, and beyond.

 \begin{acknowledgments}
We thank Rui Zhang, Noé Atzin, and Ali Mozaffari for many fruitful and enjoyable discussions. 
We also thank Amin Haeri and S.J. Kole for their valuable comments on the manuscript. 
We are especially grateful to Hans Christian Öttinger for his guidance in developing the theory.

J.S.H. also acknowledges support from the UChicago MRSEC Graduate Fellowship, NSF award number DMR-2011854.

\end{acknowledgments}

\section*{Author Contributions}
J.S.H. proposed and executed the study: developed the theoretical framework, co-developed the computational methodology, performed and analyzed the simulations, and wrote the manuscript.
P.Z.R. co-developed the computational methodology.
J.d.P. conceived the project, provided resources and supervision, and contributed to the discussion of results.
All authors contributed to the final review.

\appendix
\renewcommand{\thefigure}{\Alph{section}.\arabic{figure}}
\setcounter{figure}{0}

\section{The model for Multi-component Liquid Crystals}
\label{Multicomponent}

Lyotropic LC experiments of interest often involve more than two components, such as a secondary LC, oil, surfactant, ions, molecular motors \cite{guillamat2016control, weirich2019self}, and a solvent. To address this, we present here a multi-component (MC) theory in which the concentration $\phi$ can describe both an isotropic liquid and an LC.

We discuss two approaches: one that is easier to implement computationally but less thermodynamically rigorous (a naïve approach), and another that is more rigorous but requires greater computational resources (the proper thermodynamic approach). We use the same variables, abbreviations, and symbols, as in the main text, unless stated otherwise.

\subsection{A naïve approach}

Here, we summarize the basic ingredients of the model, including the energy and entropy functionals, a brief discussion of the structure of the Poisson and friction matrices, and the general form of the time-evolution equations required for simulations. We repeat the process presented in Section \ref{Model}, with the distinction that each component and its parameters are now labeled with $\Lambda$ to indicate its number. Thus, $\phi_\Lambda$ and $\bm{Q}_\Lambda$ represent the concentration and tensor order of the $\Lambda$-th component. This yields $\bm{x}^{mc} = (\rho^{mc}, \bm{m}^{mc}, \epsilon^{mc}, \phi_1, \dots, \phi_\Upsilon, \bm{Q}_1, \dots, \bm{Q}_\Upsilon)$, which we abbreviate as $\bm{x}^{mc} = (\rho^{mc}, \bm{m^{mc}}, \epsilon^{mc}, \phi_\Lambda, \bm{Q}_\Lambda)$, with \(\Upsilon\) denoting the total number of components. The superscript $mc$ indicates a multi-component variable, e.g., the total mass density becomes that of the mixture.

Next, the MC total energy $E^{mc}$ reads
 \begin{eqnarray} \label{totalE_multi}
E^{mc}(\bm{x}^{mc}) &=& \int \mathcal{E}^{mc}(\bm{x}^{mc}) \, dV =  \int \bigg\{
 \frac{1}{2} \frac{({m_{k}^{mc}})^{2} }{\rho^{mc}} + \epsilon^{mc}
 + \sum_{\Lambda=1}^\Upsilon  \frac{1}{2} \kappa_{E, \Lambda} \! \left(\frac{\partial\phi_\Lambda}{\partial r_k}\right)^2    
\nonumber \\ &+& 
 \sum_{\Lambda=1}^\Upsilon \epsilon_{\rm LdG,E}(\rho^{mc}, \phi_\Lambda,\bm{Q}_\Lambda) + \sum_{\Lambda=1}^\Upsilon  \frac{1}{2} K_\Lambda(\phi_\Lambda) \left[ \frac{\partial Q_{ij}}{\partial r_k} \frac{\partial Q_{ij}}{\partial r_k} \right]_\Lambda \, 
 \\ \nonumber
 &+& \sum_{\substack{\Lambda=1 \\ \Lambda<\Theta}}^\Upsilon \, \rho^{mc} \chi_{\Lambda\Theta} \phi_{\Lambda} \phi_{\Theta}  +   \sum_{\Lambda=1}^\Upsilon  \frac{1}{2} \kappa_\Lambda' \!
 \left[\left(\frac{\partial\phi}{\partial r_i}\right) \left( Q_{ij} + \frac{I_{ij}}{d}  \right) \left(\frac{\partial\phi}{\partial r_j}\right) \right]_\Lambda 
 \\ \nonumber
 &+& \sum_{\substack{\Lambda=1 \\ \Lambda<\Theta}}^\Upsilon \,  W_{R, \Lambda\Theta} \bigg(Q_{ij,\Lambda} - Q_{ij,\Theta} \bigg)^2  \bigg\} dV ,
 \end{eqnarray}
where some energies are summed over $\Lambda$. In this equation and in what follows, this notation is used systematically: the index $\Lambda$, when applied in a function, tags all terms associated with $\phi$ and $\bm{Q}$. The parameters \(\chi\) and \(W_R\), and their corresponding summation, are also parameterized by a second index \(\Theta\), representing pairwise interactions with a second liquid. We also include an energetic contribution inspired by the Rapini-Papoular potential \cite{rapini1969distorsion}, the last term in Eq. (\ref{totalE_multi}), which acts as a harmonic term that penalizes differences between the orientations of different LCs in contact (\(\Lambda\) and \(\Theta\)), aiding their alignment. The parameter \(W_R\) controls the alignment strength. Other energetic contributions are already described in Section \ref{Model}.

Analogously, the MC entropy $S^{mc}$ prays
\begin{eqnarray} \label{totalS_multi}
S^{mc}(\bm{x}^{mc}) &=& \int  \bigg[s_0(\rho^{mc}, \epsilon^{mc}) + \mathcal{S}^{mc}(\bm{x}^{mc})  \bigg] dV 
\nonumber \\ &=& \int \bigg[
s_0( \rho^{mc}, \epsilon^{mc}) + \sum_{\Lambda=1}^\Upsilon  \frac{1}{2} \kappa_{S,\Lambda} \! \left(\frac{\partial\phi_{\Lambda}}{\partial r_k}\right)^2 
- \rho^{mc} \sum_{\Lambda=1}^\Upsilon c_\Lambda \phi_\Lambda \ln \phi_\Lambda  
\nonumber \\ 
&+& \sum_{\Lambda=1}^\Upsilon \epsilon_{\rm LdG,S}(\rho^{mc}, \phi_\Lambda, \bm{Q}_\Lambda) 
\bigg] dV,
\end{eqnarray}
with the third term in the second line generalizing the entropy of mixing, $ s_{\rm mix}$.

Furthermore, the MC Poisson matrix $\bm{L}^{mc}$  takes the form:
\begin{equation}\label{LCPoisson_multi}
\bm{L}^{mc} = - \left(
\begin{array}{ccccc}
0 & \frac{\partial}{\partial r_k} \rho^{mc} & 0 & 0 & 0 \\
\rho^{mc} \frac{\partial}{\partial r_i}
& \frac{\partial}{\partial r_k} m_i^{mc} + m_k^{mc} \frac{\partial}{\partial r_i}
& \epsilon^{mc} \frac{\partial}{\partial r_i} + \frac{\partial}{\partial r_m} \pi^{S(mc)}_{mi}
& - \frac{\partial \phi_\Lambda,}{\partial r_i} & \bm{L}_{25,\Lambda} \\
0 & \frac{\partial}{\partial r_k} \epsilon^{mc} + \pi^{S(mc)}_{km} \frac{\partial}{\partial r_m}
& 0 & 0 & 0 \\
0 & \frac{\partial \phi_\Lambda,}{\partial r_k} & 0 & 0 & 0 \\
0 & \bm{L}_{52,\Lambda} & 0 & 0 & 0 \\
\end{array} \right) ,
\end{equation}

where we present a shortened version, with some entries omitted (normally indicated by dots $\dots$), similar to the vector $\bm{x}^{mc}$, while assuming that each component is convected with a common velocity ($\bm{v}^{mc} = \bm{m}^{mc} / \rho^{mc}$). The entries $\bm{L}_{25,\Lambda}$ and $\bm{L}_{52,\Lambda}$ correspond to the labeled versions of Eqs. (\ref{L25expr}) and (\ref{L52expr}). Moreover, $\bm{\pi}^{S(mc)}$ is presented when introducing the momentum balance. 

The MC friction matrix $\bm{M}^{mc}$ incorporates the sum over all components of the physical mechanisms considered in Section \ref{Model} (relaxation, diffusion, and slip processes):
\begin{equation}\label{Mcont_mc}
\bm{M}^{mc} = 
\sum_{\Lambda=1}^{\Upsilon} (\bm{M}^{\rm diff}_{\phi})_{_\Lambda} 
+ \sum_{\Lambda=1}^{\Upsilon} ( \bm{M}^{\rm relax}_{\bm{Q}})_\Lambda 
+ \sum_{\Lambda=1}^{\Upsilon} (\bm{M}^{\rm slip}_{\bm{Q}})_\Lambda,
\end{equation}
where the matrices correspond to the labeled versions of Eqs. (\ref{Mdif}), (\ref{MrelQ}), and (\ref{slip_matrix}), using $E^{mc}$, $\rho^{mc}$, and $1/T^{mc} = \partial s_0(\rho^{mc}, \epsilon^{mc}) / \partial \epsilon^{mc}$, as well as the labeled parameters $\bm{R}_\Lambda$, $\lambda_\Lambda$, and $\bm{D}_\Lambda$, the latter specifying an average diffusion tensor for the $\Lambda$-th component in the mixture.

The MC LC time-evolution equations are thus given by
\begin{equation}\label{evoleqrho_multi}
\frac{\partial\rho^{mc}}{\partial t} = - \frac{\partial}{\partial r_k} (\rho^{mc} \, v^{mc}_k)  ,
\end{equation}
\begin{equation}\label{evoleqM_multi}
\frac{\partial m_i^{mc}}{\partial t} = -\frac{\partial}{\partial r_m} \bigg[v_m^{mc} m_i^{mc} + \pi_{mi}^{S(mc)} + \pi_{mi}^{E(mc)} - \sum_{\Lambda=1}^\Upsilon \pi_{mi, \Lambda}^{slip} \bigg] ,
\end{equation}
\begin{eqnarray}\label{evoleqeps_multi}
&&\frac{\partial\epsilon^{mc}}{\partial t} = -\frac{\partial}{\partial r_k} \left( v_k^{mc} \, \epsilon^{mc}\right) - \pi^{S(mc)}_{km} \, W_{km}^{mc} 
\\ \nonumber
&+&  \sum_{\Lambda=1}^\Upsilon \bigg\{ \frac{\delta E^{mc}}{\delta Q_{kl}} R_{\Lambda} \left( \frac{1}{T^{mc}} \frac{\delta E^{mc}}{\delta Q_{kl}} - \frac{\delta S^{mc}}{\delta Q_{kl}} \right) 
\\ \nonumber
&+& T^{mc}\bigg(\lambda_\Lambda - 1 \bigg) \bigg[ 2 \left( Q_{ij} + \frac{I_{ij}}{d} \right) \frac{\delta S^{mc}}{\delta Q_{ij}} \left( Q_{lk} + \frac{I_{lk}}{d} \right) W_{lk}^{mc} 
\\ \nonumber
&-& \left( Q_{ik} + \frac{I_{ik}}{d} \right)  W_{kj}^{mc} \, \frac{\delta S^{mc}}{\delta Q_{ij}}  - \left( Q_{ik} + \frac{I_{ik}}{d} \right)  W_{jk}^{mc} \, \frac{\delta S^{mc}}{\delta Q_{ij}} \bigg]
\\ \nonumber
&-&  \frac{\delta E^{mc}}{\delta \phi} \frac{1}{\rho^{mc}} \frac{\partial}{\partial r_k} 
\rho^{mc} D_{kl, \Lambda} \, \frac{\partial}{\partial r_l} \frac{1}{\rho^{mc}} \,
\left( \frac{1}{T^{mc}} \frac{\delta E^{mc}}{\delta \phi} - \frac{\delta S^{mc}}{\delta \phi} \right) \bigg\}_\Lambda ,
\end{eqnarray}
\begin{equation}\label{evoleqphi_multi}
\frac{\partial\phi_\Lambda}{\partial t} = - v_k^{mc} \, \frac{\partial \phi_\Lambda}{\partial r_k} 
+  \frac{1}{\rho^{mc}} \frac{\partial}{\partial r_k} 
\rho^{mc} D_{kl, \Lambda} \, \frac{\partial}{\partial r_l} \frac{1}{\rho^{mc}} \,
\left( \frac{1}{T^{mc}} \frac{\delta E}{\delta \phi_\Lambda} - \frac{\delta S}{\delta \phi_\Lambda} \right)  ,
\end{equation}
\begin{eqnarray}
&&\frac{\partial Q_{ij, \Lambda}}{\partial t}= - v_k^{mc} \, \frac{\partial Q_{ij, \Lambda}}{\partial r_k} 
+ \left( \lambda_\Lambda\Xi_{ik}^{mc} + \Omega_{ik}^{mc}  \right) \left( Q_{kj, \Lambda} + \frac{I_{kj}}{d}  \right)  
\\ \nonumber
&&+ \left( Q_{ik, \Lambda}  + \frac{I_{ik}}{d} \right) \left( \lambda_\Lambda \Xi_{kj}^{mc} - \Omega_{kj}^{mc}  \right)  
- 2\lambda_\Lambda\left(  Q_{ij, \Lambda}  + \frac{I_{ij}}{d} \right) \left(  Q_{kl, \Lambda} + \frac{I_{kl}}{d} \right) W_{lk}^{mc} 
\\ \nonumber
&&- R_\Lambda \left( \frac{1}{T^{mc}} \frac{\delta E}{\delta Q_{ij, \Lambda}}
- \frac{\delta S}{\delta Q_{ij_\Lambda}} \right) ,
\label{evoleqQ_multi}
\end{eqnarray}
with
\begin{eqnarray}\label{energystressfull_multi}
\bm{\pi}_{mi}^{E(mc)} &=&  -\, \bigg\{
\bigg[ \rho^{mc} \frac{\partial}{\partial r_l} \left( \frac{\partial \mathcal{E}^{mc}}{\partial (\partial_l \rho^{mc})} \right)  + 
\epsilon^{mc} \frac{\partial}{\partial r_l} \left( \frac{\partial \mathcal{E}^{mc}}{\partial (\partial_l \epsilon^{mc})} \right)
- \rho^{mc} \frac{\partial \mathcal{E}^{mc}}{\partial \rho^{mc}}
\\ \nonumber
&-& \epsilon^{mc} \frac{\partial \mathcal{E}^{mc}}{\partial \epsilon^{mc}} + \mathcal{E}^{mc}  \bigg]  I_{mi}  - \bigg[ \left( \frac{\partial \rho^{mc}}{\partial r_i} \right) \left(\frac{\partial \mathcal{E}^{mc}}{\partial (\partial_m \rho^{mc})} \right)  
+ \left( \frac{\partial \epsilon^{mc}}{\partial r_i} \right) \left(\frac{\partial \mathcal{E}^{mc}}{\partial (\partial_m \epsilon^{mc})} \right)  
\\ \nonumber
&+& \sum_{\Lambda=1}^\Upsilon \left( \frac{\partial \phi_\Lambda}{\partial r_i} \right) \left(\frac{\partial \mathcal{E}^{mc}}{\partial (\partial_m \phi_\Lambda)} \right)
    + \sum_{\Lambda=1}^\Upsilon \left[   \left( \frac{\partial Q_{kl}}{\partial r_i} \right) \left(\frac{\partial \mathcal{E}^{mc}}{\partial (\partial_m Q_{kl})} \right) \right]_\Lambda \bigg] 
  \\ \nonumber
&-& 2 \sum_{\Lambda=1}^\Upsilon  \bigg[ \left(Q_{mi} + \frac{I_{mi}}{d}\right) \left(Q_{kl} + \frac{I_{kl}}{d}\right) \frac{\delta E^{mc}}{\delta Q_{kl}} - \left( Q_{ml} + \frac{I_{ml}}{d} \right) \frac{\delta E^{mc}}{\delta Q_{li}} \bigg]_\Lambda
\bigg\} ,
\end{eqnarray}
and
\begin{eqnarray}\label{entropystressfull_multi}
\bm{\pi}_{mi}^{S(mc)} &=&  T^{mc}\, \bigg\{
\bigg[ \rho^{mc} \frac{\partial}{\partial r_l} \left( \frac{\partial \mathcal{S}^{mc}}{\partial (\partial_l \rho^{mc})} \right)  + 
\epsilon^{mc} \frac{\partial}{\partial r_l} \left( \frac{\partial \mathcal{S}^{mc}}{\partial (\partial_l \epsilon^{mc})} \right)
- \rho^{mc} \frac{\partial \mathcal{S}^{mc}}{\partial \rho^{mc}} 
\\ \nonumber
&-& \epsilon^{mc} \frac{\partial \mathcal{S}^{mc}}{\partial \epsilon^{mc}} 
+ \mathcal{S}^{mc} + \frac{p_0^{mc}}{T^{mc}} \bigg]  I_{mi} 
- \bigg[ \left( \frac{\partial \rho^{mc}}{\partial r_i} \right) \left(\frac{\partial \mathcal{S}^{mc}}{\partial (\partial_m \rho^{mc})} \right)  
\\ \nonumber
&+& \left( \frac{\partial \epsilon^{mc}}{\partial r_i} \right) \left(\frac{\partial \mathcal{S}^{mc}}{\partial (\partial_m \epsilon^{mc})} \right)  + \sum_{\Lambda=1}^\Upsilon \left( \frac{\partial \phi_\Lambda}{\partial r_i} \right) \left(\frac{\partial \mathcal{S}^{mc}}{\partial (\partial_m \phi_\Lambda)} \right) 
  \\ \nonumber
&+& \sum_{\Lambda=1}^\Upsilon \left[   \left( \frac{\partial Q_{kl}}{\partial r_i} \right) \left(\frac{\partial \mathcal{S}^{mc}}{\partial (\partial_m Q_{kl})} \right) \right]_\Lambda  \bigg]
\\ \nonumber
&-& 2 \sum_{\Lambda=1}^\Upsilon  \bigg[ \left(Q_{mi} + \frac{I_{mi}}{d}\right) \left(Q_{kl} + \frac{I_{kl}}{d}\right) \frac{\delta S^{mc}}{\delta Q_{kl}} 
- \left( Q_{ml} + \frac{I_{ml}}{d} \right) \frac{\delta S^{mc}}{\delta Q_{li}} \bigg]_\Lambda
\bigg\} .
\end{eqnarray}

Note that $p^{mc}_0$ and $\pi_{mi, \Lambda}^{slip}$ are the indexed forms of Eqs.~(\ref{p}) and (\ref{slip_stress}), and, of course, $W_{ij}^{mc} = \partial v_i^{mc} / \partial r_j$,
$\Xi_{jk}^{mc} = (W_{jk}^{mc}+ W_{kj}^{mc})/2$, and
$\Omega_{jk}^{mc} = (W_{jk}^{mc}- W_{kj}^{mc})/2$. In addition, the equations are subject to the mass-conservation constraint:
\begin{equation}\label{concentration_equals_one}
\sum_{\Lambda=1}^{\Upsilon} \phi_\Lambda = 1,
\end{equation}
which transforms our PDEs into a differential-algebraic system, for which numerical methods exist (in Julia, they are solved by calling a \texttt{DAEProblem} of the \texttt{DifferentialEquations.jl} package). However, it is also possible to incorporate such restriction directly into the dynamical PDEs, rather than enforcing it via an algebraic equation, as described in the following subsection.

\subsection{Constraints in a diffusion system}

Noting that diffusion is a process thought to occur between two components, in deriving our MC equations, we observe that Eq.~(\ref{evoleqphi_multi}) utilizes an average diffusion tensor ($\bm{D}_\Lambda$) rather than a pairwise tensor between components (\(\Lambda\) and \(\Theta\)). This prevents us from verifying that the diffusive flux of \(\Lambda\) into \(\Theta\) equals the opposite flux of \(\Theta\) into \(\Lambda\), as only the average flux of \(\Lambda\) into the remaining mixture can be computed, leading to a loss of symmetry and a well-defined thermodynamic structure.

To address this, we proceed by analogy with the reasoning in \cite{ottinger09constraints}, which indicates that our focus should be on the first sum of Eq.~(\ref{Mcont_mc}), as it is the only term dependent on $\phi$. There, it is also proved that the MC Poisson operator remains unmodified. The issue arises because the entries of $(\bm{M}^{\rm diff}_{\phi})_{\Lambda}$ are associated with the pair $\phi_\Lambda \phi_\Lambda$ rather than with $\phi_\Lambda \phi_\Theta$, which would in turn give rise to pairwise diffusive fluxes between components. Furthermore, incorporating the constraint in Eq.~(\ref{concentration_equals_one}) into the MC friction matrix leads to:
\begin{equation}\label{M_mc_restricted}
\bm{M}^{mc} = 
\sum_{\Lambda,\Theta=1}^{\Upsilon}\bm{M}^{\rm diff-pair}_{\phi}
+ \sum_{\Lambda=1}^{\Upsilon} ( \bm{M}^{\rm relax}_{\bm{Q}})_\Lambda 
+ \sum_{\Lambda=1}^{\Upsilon} (\bm{M}^{\rm slip}_{\bm{Q}})_\Lambda,
\end{equation}
with
\begin{equation}\label{Mdif-pair}
\bm{M}^{\rm diff-pair}_{\phi} = \left( \begin{matrix}
- \frac{\delta E^{mc}}{\delta \phi_{\Lambda}} \frac{1}{\rho^{mc}} \frac{\partial}{\partial \bm{r}}
\cdot \rho^{mc} D_{\Lambda\Theta}  \frac{\partial}{\partial \bm{r}} \frac{1}{\rho^{mc}} \frac{\delta E^{mc}}{\delta \phi_{\Theta}}
& \frac{\delta E^{mc}}{\delta \phi_{\Lambda}} \frac{1}{\rho^{mc}} \frac{\partial}{\partial \bm{r}}
\cdot \rho^{mc} D_{\Lambda\Theta}  \frac{\partial}{\partial \bm{r}} \frac{1}{\rho^{mc}} \\ 
\frac{1}{\rho^{mc}} \frac{\partial}{\partial \bm{r}} \cdot \rho^{mc} D_{\Lambda\Theta} 
\frac{\partial}{\partial \bm{r}} \frac{1}{\rho^{mc}} \frac{\delta E^{mc}}{\delta \phi_{\Theta}}
& - \frac{1}{\rho^{mc}} \frac{\partial}{\partial \bm{r}} \cdot \rho^{mc} D_{\Lambda\Theta}  \frac{\partial}{\partial \bm{r}} \frac{1}{\rho^{mc}}
\end{matrix} \right) ,
\end{equation}
which represents the block of $\bm{M}^{\rm diff}_{\phi}$ (Eq.~\ref{Mdif}) associated with $\epsilon^{mc}$ and $\phi$, now accounting for pairwise interactions. Its entries are related to the variables as follows: the subscript of $\phi$ is $\Lambda$ when it appears in the row (first matrix index), for instance, the (2,1)-element corresponds to $(\phi_\Lambda, \epsilon^{mc})$, and $\Theta$ when it appears in the column (second matrix index); for example, the (2,2)-element corresponds to the $(\phi_\Lambda, \phi_\Theta)$ coupling.

For simplicity, we take the diffusion parameter to be the symmetric and positive-semidefinite matrix $D_{\Lambda\Theta}$, in place of a matrix whose entries are anisotropic tensors, which satisfies the property:
\begin{equation}\label{D_conditions}
\sum_{\Lambda=1}^{\Upsilon} D_{\Lambda \Theta} = \sum_{\Theta=1}^{\Upsilon} D_{\Lambda \Theta} = 0,
\end{equation}
an additional constraint that ensures the sum of concentrations equals 1, and that the diffusion fluxes between each pair of components have the same magnitude but opposite direction. Equation~\eqref{M_mc_restricted} then modifies the energy balance and the diffusion equation of $\phi_\Lambda$ as:
\begin{eqnarray}\label{evoleqeps_multi}
&&\frac{\partial\epsilon^{mc}}{\partial t} = -\frac{\partial}{\partial r_k} \left( v_k^{mc} \, \epsilon^{mc}\right) - \pi^{S(mc)}_{km} \, W_{km}^{mc} 
\\ \nonumber
&+&  \sum_{\Lambda=1}^\Upsilon \bigg\{ \frac{\delta E^{mc}}{\delta Q_{kl}} R_{\Lambda} \left( \frac{1}{T^{mc}} \frac{\delta E^{mc}}{\delta Q_{kl}} - \frac{\delta S^{mc}}{\delta Q_{kl}} \right) 
\\ \nonumber
&+& T^{mc}\bigg(\lambda_\Lambda - 1 \bigg) \bigg[ 2 \left( Q_{ij} + \frac{I_{ij}}{d} \right) \frac{\delta S^{mc}}{\delta Q_{ij}} \left( Q_{lk} + \frac{I_{lk}}{d} \right) W_{lk}^{mc} 
\\ \nonumber
&-& \left( Q_{ik} + \frac{I_{ik}}{d} \right)  W_{kj}^{mc} \, \frac{\delta S^{mc}}{\delta Q_{ij}}  - \left( Q_{ik} + \frac{I_{ik}}{d} \right)  W_{jk}^{mc} \, \frac{\delta S^{mc}}{\delta Q_{ij}} \bigg] \bigg\}_\Lambda
\\ \nonumber
&-& \sum_{\Lambda, \Theta=1}^\Upsilon \frac{\delta E^{mc}}{\delta \phi_{\Lambda}} \frac{1}{\rho^{mc}} \frac{\partial}{\partial r_k} 
\rho^{mc} D_{\Lambda\Theta} \, \frac{\partial}{\partial r_l} \frac{1}{\rho^{mc}} \,
\left( \frac{1}{T^{mc}} \frac{\delta E^{mc}}{\delta \phi_{\Theta}} - \frac{\delta S^{mc}}{\delta \phi_{\Theta}} \right)  ,
\end{eqnarray}
and 
\begin{equation}\label{evoleqphi_multi_restri}
\frac{\partial\phi_\Lambda}{\partial t} = - v_k^{mc} \, \frac{\partial \phi_\Lambda}{\partial r_k} 
+  \sum_{\Theta=1}^{\Upsilon}\frac{1}{\rho^{mc}} \frac{\partial}{\partial r_k} 
\rho^{mc} D_{\Lambda\Theta} \, \frac{\partial}{\partial r_l} \frac{1}{\rho^{mc}} \,
\left( \frac{1}{T^{mc}} \frac{\delta E}{\delta \phi_\Theta} - \frac{\delta S}{\delta \phi_\Theta} \right)  .
\end{equation}

Note that these two equations, together with Eq. (\ref{D_conditions}), may be more computationally challenging, as they require additional Laplacians for each component of the system (the last sums on their RHS), but still preserve thermodynamic consistency. A similar pairwise sum could, in principle, also be included in the evolution of $\bm{Q}_{\Lambda}$ by introducing a relaxation parameter $\bm{R}_{\Lambda\Theta}$. However, we assume the reasonable approximation $\bm{R}_{\Lambda\Theta} = \bm{R} \, \bm{I}_{\Lambda\Theta}$, which implies that, after summation, only the diagonal terms $\bm{R}_{\Lambda\Lambda} = \bm{R}_\Lambda$ remain, as used in our model, since $\bm{Q}$ is not, to the best of our knowledge, expected to follow any restriction with other components (only with itself), i.e., scalar order parameters are not conserved quantities.

To conclude, it is important to emphasize that the Newtonian stress tensor has not been included in the energy balance yet. Additional physical effects can be incorporated by modifying the energy or entropy functionals, or by changing the $\bm{M}$ matrix. To further simplify the equations, one could consider that $E_{free} = E - T \ S$, where $E_{free}$ is the Helmholtz free energy. We also note that none of our pressure contributions in the momentum balance—binary or MC—include a term of the form $\phi \, \frac{\delta E}{\delta \phi}$, or $\phi \, \frac{\delta S}{\delta \phi}$ as the mass fraction, lacking a volumetric ratio or density-like character, does not contribute mechanically to the stress. The absence of these terms facilitates numerical implementation.

\section{From Poisson integrals to matrix notation}
\label{BE:appendix}

Here, we show how to translate the Poisson bracket in its integral form into the corresponding entry of the $\bm{L}$ matrix. We use the same variables, abbreviations, symbols and conventions as in the main text, unless otherwise noted.

Specifically, Eq. (11.5-37) in \cite{BerisEdwards} presents the BE Poisson bracket for $\bm{Q}$ , denoted by the braces $\{\cdot\ , \cdot\}$, and reads:
\begin{eqnarray}\label{BEPoissonintegral}
    \nonumber
    \{A_F, A_G\} = &-& \int m_i \left[ \frac{\delta A_F}{\delta m_k} \frac{\partial}{\partial r_k} \frac{\delta A_G}{\delta m_i} -  \frac{\delta A_G}{\delta m_k} \frac{\partial}{\partial r_k} \frac{\delta A_F}{\delta m_i}\right] dV  \\ \nonumber
    &-& \int Q'_{ij} \, \frac{\partial}{\partial r_k} \left[ \frac{\delta A_F}{\delta m_k} \left(  \frac{\delta A_G}{\delta Q'_{ij}} \right)  -  \frac{\delta A_G}{\delta m_k} \left(  \frac{\delta A_F}{\delta Q'_{ij}} \right) \right] dV  \\ \nonumber
    &-& \int Q'_{ik} \left[ \left(\frac{\partial }{\partial r_k} \frac{\delta A_F}{\delta m_j} \right) \frac{\delta A_G}{\delta Q'_{ij}}  - \left( \frac{\partial}{\partial r_k} \frac{\delta A_G}{\delta m_j} \right)\frac{\delta A_F}{\delta Q'_{ij}}  \right] dV  \\ \nonumber
    &-& \int Q'_{jk} \left[ \left(\frac{\partial }{\partial r_k} \frac{\delta A_F}{\delta m_i} \right) \frac{\delta A_G}{\delta Q'_{ij}}  - \left( \frac{\partial}{\partial r_k} \frac{\delta A_G}{\delta m_i} \right)\frac{\delta A_F}{\delta Q'_{ij}}  \right] dV \\ 
    &+& 2 \int Q'_{ij}Q'_{kl} \left[ \left(\frac{\partial }{\partial r_j} \frac{\delta A_F}{\delta m_i} \right) \frac{\delta A_G}{\delta Q'_{kl}}  - \left( \frac{\partial}{\partial r_j} \frac{\delta A_G}{\delta m_i} \right)\frac{\delta A_F}{\delta Q'_{kl}}  \right] dV,
\end{eqnarray}
where $\bm{Q}' = \bm{Q} + \bm{I}/d$, implying $ \delta / \delta \bm{Q}' = \delta / \delta \bm{Q} $, and $A_F$ and $A_G$ are auxiliary functionals. 

In the original BE formulation, the bracket is expressed in terms of $\bm{v}$, assuming incompressibility. To adapt it to our framework, we use $\bm{m}$ instead and replace only its second line with the corresponding expression from GENERIC (the first line of Eq. (2.92) in \cite{ottinger2005beyond}). These minor adjustments allow us to generalize the BE bracket to compressible fluids while preserving its main structure.

For our purposes, the relevant terms are those in lines 2 through 5, as the first line only describes momentum-momentum coupling (see their functional derivatives). The second line resembles a convective bracket for scalar-like, or non-volumetric, variables, like $\phi$. The third and fourth correspond to additional contributions arising from the two extra indices in the tensor, and the fifth enforces the traceless condition.

The key relationship connecting this bracket to $\bm{L}$ is given by Eq. (1.6) of \cite{ottinger2005beyond}:
\begin{equation}\label{LmatrixPoissonbrackets}
    \{A_F, A_G\} = \frac{\delta A_F}{\delta \bm{x}} \cdot \bm{L}(\bm{x}) \cdot  \frac{\delta A_G}{\delta \bm{x}}.
\end{equation}
To find the $\bm{L}_{25}$ and $\bm{L}_{52}$ entries, associated with $\bm{Q}$ and $\bm{m}$, we separate the terms that pair $\delta A_{F} / \delta \bm{m}$ with $\delta A_G / \delta \bm{Q}'$ from those that combine $\delta A_{G} / \delta \bm{m}$ with $\delta A_F / \delta \bm{Q}'$, and rewrite Eq. (\ref{LmatrixPoissonbrackets}), using Eq. (\ref{BEPoissonintegral}), as:
\begin{eqnarray}\label{L25_derivation}
    \nonumber
    &&\{A_F, A_G\}_{\bm{m}, \bm{Q}'} = \frac{\delta A_F}{\delta \bm{m}} \cdot \bm{L}_{25} \cdot  \frac{\delta A_G}{\delta \bm{Q}'}
    \\ \nonumber
    &=&  - \int  \bigg[  Q'_{ij} \, \frac{\partial}{\partial r_k} \left( \frac{\delta A_F}{\delta m_k}   \frac{\delta A_G}{\delta Q'_{ij}}    \right)+  Q'_{ik} \left(\frac{\partial }{\partial r_k} \frac{\delta A_F}{\delta m_j} \right) \frac{\delta A_G}{\delta Q'_{ij}} \\ \nonumber
    &+& Q'_{jk}  \left(\frac{\partial }{\partial r_k} \frac{\delta A_F}{\delta m_i} \right) \frac{\delta A_G}{\delta Q'_{ij}}  -  2  Q'_{ij}Q'_{kl}  \left(\frac{\partial }{\partial r_j} \frac{\delta A_F}{\delta m_i} \right) \frac{\delta A_G}{\delta Q'_{kl}}   \bigg]  dV 
    \\ 
    &=&  \int\frac{\delta A_F}{\delta m_k} \bigg[ \left( \frac{\partial Q'_{ij}}{\partial r_k} \right) + \frac{\partial}{\partial r_m} I_{jk} \, Q'_{im}   \, +  \frac{\partial}{\partial r_m} I_{ik} \, Q'_{jm} - 2\frac{\partial}{\partial r_m} Q'_{km} Q'_{ij}\bigg] \frac{\delta A_G}{\delta Q'_{ij}} dV,
\end{eqnarray}
and similarly as:
\begin{eqnarray}\label{L52_derivation}
    \nonumber
    &&\{A_F, A_G\}_{\bm{Q}', \bm{m}} = \frac{\delta A_F}{\delta \bm{Q}'} \cdot \bm{L}_{25} \cdot  \frac{\delta A_G}{\delta \bm{m}}
    \\ \nonumber
    &=&  \int  \bigg[  Q'_{ij} \, \frac{\partial}{\partial r_k} \left( \frac{\delta A_G}{\delta m_k}   \frac{\delta A_F}{\delta Q'_{ij}}    \right)+  Q'_{ik} \left(\frac{\partial }{\partial r_k} \frac{\delta A_G}{\delta m_j} \right) \frac{\delta A_F}{\delta Q'_{ij}} \\ \nonumber
    &+& Q'_{jk}  \left(\frac{\partial }{\partial r_k} \frac{\delta A_G}{\delta m_i} \right) \frac{\delta A_F}{\delta Q'_{ij}}  -  2  Q'_{ij}Q'_{kl}  \left(\frac{\partial }{\partial r_j} \frac{\delta A_G}{\delta m_i} \right) \frac{\delta A_F}{\delta Q'_{kl}}   \bigg]  dV 
    \\ 
    &=&   \int\frac{\delta A_F}{\delta Q'_{ij}} \bigg[ -\left( \frac{\partial Q'_{ij}}{\partial r_k} \right) + I_{jk} \, Q'_{im} \frac{\partial}{\partial r_m}    \, + I_{ik} \, Q'_{jm} \frac{\partial}{\partial r_m}  - 2 Q'_{km} Q'_{ij} \frac{\partial}{\partial r_m} \bigg] \frac{\delta A_G}{\delta m_k} dV,
\end{eqnarray}
where we have swapped some indices, either by introducing a new index $m$ or by using the identity matrix, to rearrange the functional derivatives so that they appear  at the ends of the integrals.

Additionally, to move some functional derivatives across the spatial derivatives whenever convenient, we apply Gauss' theorem, which is expressed here as:
\begin{equation}\label{divergence_theorem}
  \int \bm{A}_F \, \cdot \frac{\partial \bm{A}_G}{\partial \bm{r}}  \,   dV   = -\int \bm{A}_G \, \left( \frac{\partial}{\partial \bm{r}} \cdot \bm{A}_F \right) \, dV,
\end{equation}
with $\bm{A}_F$ and $\bm{A}_G$ denoting auxiliary tensor functions. Note that, as usual, we neglect the influence of boundary terms (e.g. $\int \bm{A}_F \bm{A}_g \, \cdot \bm{\hat{n}} \ dS = \bm{0}$, where $dS$ indicates the surface element and $\bm{\hat{n}}$ is the unit normal vector), by assuming that the functions vanish sufficiently at the boundaries. 

Also observe that whether the required functional derivative (on the right) lies outside or inside the spatial derivative determines whether this spatial derivative acts only on the terms in the numerator or on the entire expression after multiplication. In a similar spirit, this is why we modify the second line of the BE Poisson bracket: if kept as it is, then after multiplication with the corresponding energy functional entry (in this illustrative case, $\bm{v}$), it yields a term of the form $\partial /\partial \bm{r} \, \cdot ( \bm{v} \bm{Q} )$, which only upon invoking incompressibility ($\partial v_i /\partial r_i   = 0$) leads to the term present in our equations, namely $\bm{v} \cdot \partial \bm{Q} / \partial \bm{r}$. 

Finally, the terms sandwiched between the functional derivatives in the last lines of Eqs. (\ref{L25_derivation}) and (\ref{L52_derivation}) correspond to the Poisson entries shown in Eqs. (\ref{L25expr}) and (\ref{L52expr}), as desired.

\section{Calculation of the entropic pressure tensor}
\label{appendix:b}

In this section, we present the derivation of equation (\ref{entstressfull}). We use the same variables, abbreviations, and symbols, as in the main text, unless stated otherwise. Given a general entropy density for a lyotropic LC mixture of the form
\begin{equation}\label{s_total_appen}
    s_{total} =s(\rho, \epsilon, \phi, \bm{Q}, \partial_{\bm{r}} \rho,  \partial_{\bm{r}} \epsilon, \partial_{\bm{r}} \phi, \partial_{\bm{r}} \bm{Q}) + s_0(\rho, \epsilon),
\end{equation}
and the corresponding total entropy given by
\begin{align} \label{total_S_appen}
S_{Total} &= \int  s_{total}(\rho, \epsilon, \phi, \bm{Q}, \partial_{\bm{r}} \rho,  \partial_{\bm{r}} \epsilon, \partial_{\bm{r}} \phi, \partial_{\bm{r}} \bm{Q}) \  dV ,
\end{align}
where $s_0$ is again the background fluid entropy and $s$ is any additional entropy in terms of the variables $\bm{x}$ and their gradients. Here we explicitly state that $s_{total}$ is not a function of $\bm{m}$ or the gradients of $\bm{m}$ and depends only on first-order derivatives of $\bm{x}$, which implies the functional derivative is
\begin{equation}\label{functional_derivative}
    \frac{\delta S_{Total}}{\delta \bm{x}} = \frac{\partial s_{total}}{\partial \bm{x}} - \frac{\partial}{\partial r_m} \left( \frac{\partial s_{total}}{\partial (\partial_m \bm{x})} \right).
\end{equation}
Furthermore, we can decompose the regular spatial gradient of equation (\ref{s_total_appen}) in terms of partial derivatives as follows
\begin{equation}\label{spatial_gradient_s}
\frac{\partial s_{total}}{\partial r_i} = \frac{\partial s}{\partial r_i} + \frac{\partial s_0}{\partial r_i}.
\end{equation}
The partial derivatives in Eq. (\ref{spatial_gradient_s}) are computed using the chain rule:
\begin{eqnarray}\label{chain_rule_s}
    \frac{\partial s}{\partial r_i} &=& \frac{\partial \rho}{\partial r_i}\frac{\partial s}{\partial \rho} + \frac{\partial^2 \rho}{\partial r_i \partial r_m} \frac{\partial s}{\partial(\partial_m \rho)} + \frac{\partial \epsilon}{\partial r_i}\frac{\partial s}{\partial \epsilon} +  \frac{\partial^2 \epsilon}{\partial r_i \partial r_m} \frac{\partial s}{\partial(\partial_m \epsilon)}  \nonumber \\ 
    &+& \frac{\partial \phi}{\partial r_i}\frac{\partial s}{\partial \phi}  + \frac{\partial^2 \phi}{\partial r_i \partial r_m} \frac{\partial s}{\partial(\partial_m \phi)} +  \frac{\partial Q_{kl}}{\partial r_i}\frac{\partial s}{\partial Q_{kl}} + \frac{\partial^2 Q_{kl}}{\partial r_i \partial r_m} \frac{\partial s}{\partial(\partial_m Q_{kl})} ,
\end{eqnarray} 
and
\begin{equation}\label{chain_rule_s_o}
   \frac{\partial s_0}{\partial r_i} = \frac{\partial \rho}{\partial r_i}\frac{\partial s_0}{\partial \rho} +  \frac{\partial \epsilon}{\partial r_i}\frac{\partial s_0}{\partial \epsilon}.
\end{equation}
As in the main text, we find the total entropy functional derivative
\begin{equation}\label{STotal_gradient_appen}
    \left( \frac{\delta S_{Total}}{\delta \bm{x}} \right)^{\dagger} = \left(  \frac{\delta S_{Total}}{\delta \rho}, \bm{0} ,  \frac{\delta S_{Total}}{\delta \epsilon},  \frac{\delta S_{Total}}{\delta \phi},  \frac{\delta S_{Total}}{\delta Q_{kl}}\right),
\end{equation}
with the superscript $\dagger$ denoting the transpose operation. 

The relationship useful for finding the entropic pressure contribution is dictated by the degeneracy condition in Eq. (\ref{eq:deg1}), which, together with Eqs. (\ref{LCPoisson}) and (\ref{STotal_gradient_appen}), reads
\begin{eqnarray}\label{degeneracy_appen}
    \bm{L} \cdot  \frac{\delta S_{Total}}{\delta \bm{x}} &=& \rho \frac{\partial}{\partial r_i} \left( \frac{\delta S_{Total}}{\delta \rho} \right) + \epsilon \frac{\partial}{\partial r_i} \left( \frac{\delta S_{Total}}{\delta \epsilon} \right) \\ \nonumber
    &+& \frac{\partial}{\partial r_m} \left[ \pi_{mi}^S  \frac{\delta S_{Total}}{\delta \epsilon} \right] - \frac{\partial \phi}{\partial r_i} \frac{\delta S_{Total}}{\delta \phi} + (L_{25})_{ikl} \frac{\delta S_{Total}}{\delta Q_{kl}} = 0_i,
\end{eqnarray}
which is the second component of the entropy degeneracy. By combining equations (\ref{s_total_appen}), (\ref{functional_derivative}), (\ref{L25expr}), and (\ref{degeneracy_appen}), we obtain
\begin{eqnarray}\label{expanded_degeneracy}
    0_i &=& \rho \frac{\partial}{\partial r_i} \left( \frac{\partial s}{\partial \rho} \right) - \rho \frac{\partial}{\partial r_i} \left[ \frac{\partial}{\partial r_m} \left( \frac{\partial s}{\partial (\partial_m \rho)} \right) \right] + \rho \frac{\partial}{\partial r_i} \left( \frac{\partial s_0}{\partial \rho} \right) \\ \nonumber
    &+& \epsilon \frac{\partial}{\partial r_i} \left( \frac{\partial s}{\partial \epsilon} \right) - \epsilon \frac{\partial}{\partial r_i} \left[ \frac{\partial}{\partial r_m} \left( \frac{\partial s}{\partial (\partial_m \epsilon)} \right) \right] + \epsilon \frac{\partial}{\partial r_i} \left( \frac{\partial s_0}{\partial \epsilon} \right)  \\ \nonumber
    &+& \frac{\partial}{\partial r_m} \left[ \pi_{mi}^S  \frac{\delta S_{Total}}{\delta \epsilon} \right] - \frac{\partial \phi}{\partial r_i} \frac{\partial s}{\partial \phi} + \frac{\partial \phi}{\partial r_i}  \frac{\partial}{\partial r_m} \left( \frac{\partial s}{\partial (\partial_m \phi)} \right) \\ \nonumber
    &-& \frac{\partial Q_{kl}}{\partial r_i} \frac{\partial s}{\partial Q_{kl}} + \frac{\partial Q_{kl}}{\partial r_i}  \frac{\partial}{\partial r_m} \left( \frac{\partial s}{\partial (\partial_m Q_{kl})} \right) + \frac{\partial}{\partial r_m} L^{\texttt{s}}_{mi} ,
\end{eqnarray}
where 
\begin{equation}\label{L^s}
     L^{\texttt{s}}_{mi} =  2 \bigg[ \left(Q_{mi} + \frac{I_{mi}}{d}\right) \left(Q_{kl} + \frac{I_{kl}}{d}\right) \frac{\delta  S_{Total}}{\delta Q_{kl}} - \left( Q_{ml} + \frac{I_{ml}}{d} \right) \frac{\delta S_{Total}}{\delta Q_{li}}  \bigg],
\end{equation}
note that in Eq. (\ref{expanded_degeneracy}), we, of course, recall $\frac{\partial s_0}{\partial (\partial_m \bm{x})} = \bm{0}$. 

To further simplify Eq. (\ref{expanded_degeneracy}), we need 1) the Leibniz product rule as
\begin{equation}\label{product_rule1}
  \bm{x}\frac{\partial}{\partial r_i} \left( \frac{\partial A_s}{\partial \bm{x}} \right) =  \frac{\partial}{\partial r_i} \left( \bm{x} \frac{\partial A_s}{\partial \bm{x}} \right) - \frac{\partial A_s}{\partial \bm{x}} \frac{\partial \bm{x}}{\partial r_i}, 
\end{equation}
and
\begin{equation}\label{product_rule2}
  \bm{x}\frac{\partial}{\partial r_i} \left[ \frac{\partial}{\partial r_m} \left( \frac{\partial A_s}{\partial (\partial_m \bm{x})} \right)    \right] =  \frac{\partial}{\partial r_i} \left[ \bm{x} \frac{\partial}{\partial r_m} \left( \frac{\partial A_s}{\partial (\partial_m \bm{x})} \right)  \right] -  \frac{\partial \bm{x}}{\partial r_i}\frac{\partial}{\partial r_m} \left( \frac{\partial A_s}{\partial (\partial_m \bm{x})} \right) , 
\end{equation}
for any arbitrary entropy density $A_s$; 2) the local-equilibrium Euler equation:
\begin{equation}\label{p_appendix}
\frac{p_0}{T} =  - \rho \frac{\partial s_0}{\partial \rho}
- \epsilon \frac{\partial s_0}{\partial \epsilon} + s_0 ,
\end{equation}
which now can be expressed, with Eq. (\ref{chain_rule_s_o}), as
\begin{equation}\label{p_derivative}
\frac{\partial}{\partial r_m}\left(-\frac{p_0}{T} \ I_{mi} \right) = \frac{\partial}{r_i} \left( \rho \frac{\partial s_0}{\partial \rho} \right)
+ \frac{\partial}{r_i} \left( \epsilon \frac{\partial s_0}{\partial \epsilon} \right) - \frac{\partial \rho}{\partial r_i}\frac{\partial s_0}{\partial \rho} -  \frac{\partial \epsilon}{\partial r_i}\frac{\partial s_0}{\partial \epsilon};
\end{equation}
and 3) the following relationship: 
\begin{eqnarray}\label{condensed_terms}
    & &\frac{\partial}{\partial r_m} \bigg[ -s\ I_{mi} + \left( \frac{\partial \rho}{\partial r_i}  \right) \left( \frac{\partial s}{\partial(\partial_m \rho)}  \right) 
    +  \left( \frac{\partial \epsilon}{\partial r_i}  \right) \left( \frac{\partial s}{\partial(\partial_m \epsilon)} \right)  \\ \nonumber
    &+& \left( \frac{\partial \phi}{\partial r_i}  \right) \left( \frac{\partial s}{\partial(\partial_m \phi)}   \right) + \left( \frac{\partial Q_{kl}}{\partial r_i}  \right) \left( \frac{\partial s}{\partial(\partial_m Q_{kl})}   \right)  \bigg] \\ \nonumber
    &=& -\frac{\partial s}{\partial \rho} \frac{\partial \rho}{\partial r_i} + \frac{\partial \rho}{\partial r_i}\frac{\partial}{\partial r_m} \left( \frac{\partial s}{\partial (\partial_m \rho)} \right)
    -\frac{\partial s}{\partial \epsilon} \frac{\partial \epsilon}{\partial r_i} + \frac{\partial \epsilon}{\partial r_i}\frac{\partial}{\partial r_m} \left( \frac{\partial s}{\partial (\partial_m \epsilon)} \right) \\ \nonumber
    &-&\frac{\partial s}{\partial \phi} \frac{\partial \phi}{\partial r_i} + \frac{\partial \phi}{\partial r_i}\frac{\partial}{\partial r_m} \left( \frac{\partial s}{\partial (\partial_m \phi)} \right) - \frac{\partial s}{\partial Q_{kl}} \frac{\partial Q_{kl}}{\partial r_i} + \frac{\partial Q_{kl}}{\partial r_i}\frac{\partial}{\partial r_m} \left( \frac{\partial s}{\partial (\partial_m Q_{kl})} \right),
\end{eqnarray}
which can be proved by expanding the derivatives in the first two lines and applying Eq. (\ref{chain_rule_s}).

Finally, inserting Equations (\ref{product_rule1}), (\ref{product_rule2}), (\ref{p_derivative}), and (\ref{condensed_terms}), into Eq. (\ref{expanded_degeneracy}) in that order, so that all terms remain inside derivatives $\frac{\partial}{\partial r_m}$, sometimes requiring a change of index from $i$ to $m$ using the identity matrix $I_{mi}$, yields the following equation:
\begin{eqnarray}\label{entropic_derivation}
   && \frac{\partial}{\partial r_m} \left[ -\rho \frac{\partial}{\partial r_k} \left( \frac{\partial s}{\partial (\partial_k \rho)} \right)  - 
 \epsilon \frac{\partial}{\partial r_k} \left( \frac{\partial s}{\partial (\partial_k \epsilon)} \right)
+ \rho \frac{\partial s}{\partial \rho} + \epsilon \frac{\partial s}{\partial \epsilon} - s - \frac{p_0}{T}  \right]  I_{mi}   \nonumber \\
&+& \frac{\partial}{\partial r_m} \bigg[ \left( \frac{\partial \rho}{\partial r_i} \right) \left(\frac{\partial s}{\partial (\partial_m \rho)} \right)  
+ \left( \frac{\partial \epsilon}{\partial r_i} \right) \left(\frac{\partial s}{\partial (\partial_m \epsilon)} \right)    \nonumber \\
&+& \left( \frac{\partial \phi}{\partial r_i} \right) \left(\frac{\partial s}{\partial (\partial_m \phi)} \right) 
+ \left( \frac{\partial Q_{kl}}{\partial r_i} \right) \left(\frac{\partial s}{\partial (\partial_m Q_{kl})} \right)   \bigg]  \nonumber \\
&+& \frac{\partial}{\partial r_m} L^{\texttt{s}}_{mi} = -\frac{\partial}{\partial r_m}\left[  \pi_{mi}^S  \frac{\delta S_{Total}}{\delta \epsilon}  \right].
\end{eqnarray}
Equation (\ref{entropic_derivation}) then proves the relationship shown in Eq. (\ref{entstressfull}), as desired.

\section{Numerical Methods}
\label{appendix:a}

In this section, we present our numerical and computational methods. We use the same variables, abbreviations, and symbols, as in the main text, unless stated otherwise. We use two distinct FD approaches for spatial discretization based on the nature of the equations being solved. For nonhydrodynamic spatial derivative terms (i.e., those that do not involve flow velocity)—such as first-order (e.g., gradient and divergence) and second-order operators (e.g., the Laplacian)—we employ a second-order accurate central difference scheme. This approach provides accurate approximations of these derivatives by using values from neighboring grid points on both sides of each axis in a discretized mesh \cite{balsara2013}.

For illustrative purposes, consider an arbitrary one-dimensional function $\psi(x)$ dependent solely on the $x$ direction, discretized on a structured mesh with a constant spacing step $\Delta x$.  In this context, the second-order (accurate) central difference scheme approximates spatial derivatives as follows:
%
%
\begin{equation} \label{1st-FD}
\frac{\partial\psi}{\partial x}\approx \frac{\psi(x + \Delta x) - \psi(x - \Delta x)}{2\Delta x},
\end{equation}
\begin{equation} \label{2nd-FD}
\frac{\partial^2\psi}{\partial x^2} \approx \frac{\psi(x + \Delta x) - 2\psi(x) + \psi(x - \Delta x) }{\Delta x^2},
\end{equation}
where the function values at the grid points immediately in the positive and negative directions from the central point $\psi(x)$ are denoted by $\psi(x + \Delta x)$ and $\psi(x - \Delta x)$.

Next, in the absence of hydrodynamics, time integration is performed using the \texttt{DifferentialEquations.jl} library \cite{rackauckas2017differentialequations}, whose documentation is available at \url{docs.sciml.ai/DiffEqDocs/stable/}. Here, we summarize the key features relevant to our purposes:

\texttt{DifferentialEquations.jl} is a versatile Julia library designed for solving DEs. To use this library, we first define the problem to solve, e.g., equations (\ref{evoleqphi}) and (\ref{evoleqQ}). We also specify the ICs, the time span from $0$ to $t_f$—where $t_f$ is the total number of time steps—along with other parameters such as constants in the equations, computational caches for efficiently storing information, the BCs, and any other required elements.

To solve the ODE (ordinary differential equation) problem, we apply a numerical method to integrate the problem over time. There are many integrators listed in the documentation; some popular ones include \texttt{Euler()} and \texttt{Tsit5()}, an adaptive higher-order Runge–Kutta method. The choice of integrator is based on empirical testing, prioritizing performance and numerical stability. In our main text, we employ the explicit higher-order integrator \texttt{CarpenterKennedy2N54()}. Additionally, we specify the time step of the integrator, \texttt{dt}, which controls the intervals at which the solution is computed—unless the method is time-adaptive, in which case \texttt{dt} changes with each iteration. In the main text, we refer to this \texttt{dt} as $\Delta t$. 

Conversely, for hydrodynamic simulations, where flow velocity is involved, we use a (first-order accurate) backward and forward FD scheme (also known as upwind scheme). This method improves accuracy and stability by accounting for the direction of the flow \cite{balsara2013}.

In a one-dimensional exemplary setting, the backward and forward scheme discretizes the first-order derivative of $\psi(x)$ (typically the only term affected by convective effects) based on the sign of the flow velocity $v$ that is moving the system.

For a positive flow velocity ($v > 0$):
\begin{equation} \label{positive-upwind}
\frac{\partial\psi}{\partial x} \approx \frac{\psi(x) - \psi(x - \Delta x)}{\Delta x},
\end{equation}
and for a negative flow velocity ($v < 0$):
\begin{equation} \label{negative-upwind}
\frac{\partial\psi}{\partial x} \approx \frac{\psi(x + \Delta x) - \psi(x)}{\Delta x},
\end{equation}
here, only one neighboring point is considered, unlike the two points in a second-order central difference scheme. For a more compact form, equations (\ref{positive-upwind}) and (\ref{negative-upwind}) are typically combined with the velocity as
\begin{equation} \label{BF-dif}
v\cdot \frac{\partial \psi}{\partial x} = v^+  \frac{ \psi(x) - \psi(x - \Delta x)}{\Delta x} + v^-  \frac{ \psi(x + \Delta x) - \psi(x)}{\Delta x} \equiv \mathrm{Op_x}(\psi, v),
\end{equation}
where
\begin{equation} \label{vplusminusdef}
v^+ = \max(v, 0), 
\qquad 
v^- = \min(v, 0),
\end{equation} 
$\max(v, 0)$ representing the maximum of $v$ and 0 (which is $v$ if $v > 0$ and 0 otherwise), and $\min(v, 0)$ denoting the minimum of $v$ and 0 ($v$ if $v < 0$, and 0 otherwise). This compact form combines backward and forward approximations for positive and negative velocities into a single expression, defined as the $\mathrm{Op_i}$ operator, with its subscript indicating both the velocity component and the discretized axis.

Now, for time integration, we only employ an iterative forward Euler’s method when using the backward and forward FD scheme, because, in our tests, this seemingly less sophisticated method appears more stable compared to those found in the \texttt{DifferentialEquations.jl} package. As before, the forward Euler's method is illustrated with a simple example. 

Given a differential equation of the form $\frac{\partial\psi(t)}{\partial t} = \Psi(t, \psi(t))$, one step of the Euler method from $t_a$ to $t_{a+1} = t_a + \Delta t_{Eu}$  is given by
\begin{equation} \label{ForwEuler}
\psi_{a+1} = \psi_a + \Delta t_{Eu} \cdot \Psi(t_a, \psi_a),
\end{equation}
where $\Delta t_{Eu}$ represents the time step between two iterations, the subscript $a$ denotes the time step index and $\Psi$ is a second auxiliary function. 

Furthermore, for multidimensional systems, the first-order upwind scheme in space requires modifications known as the dimensional splitting method \cite{balsara2013}. This method involves approximating the derivatives, along one dimension (e.g., $x$-direction), using equation (\ref{BF-dif}) and then updating the values of the function at each grid point based on this computation. Specifically, the old values are replaced with new ones through the Euler integrator. After the solution is updated, the procedure is then applied along the next dimension (e.g., $y$-direction) using the updated values from the previous step. In a 3D case, this procedure is followed sequentially for all three dimensions. The updated values from each dimension form the final solution, representing one single time step.

In mathematical terms, the dimensional splitting method can be understood through the discretization of the 2D advection equation
\begin{equation} \label{eq:advection}
\frac{\partial \psi}{\partial t} = -v_x \frac{\partial \psi}{\partial x} - v_y \frac{\partial \psi}{\partial y},
\end{equation}
with the update step for $\psi$ given by:
\begin{equation} \label{Ad_discre}
\psi_{a+1} = \psi_a - \Delta t_{Eu} \left[\mathrm{Op_x}(\psi_a, v_x) + \mathrm{Op_y}\left(\psi_a +\Delta t_{Eu} \cdot \mathrm{Op_x}(\psi_a, v_x), v_y\right) \right],
\end{equation}
where the subscripts in the velocity components denote the usual Cartesian directions. In equation (\ref{Ad_discre}), the advection (or movement) is first applied in the $x$-direction and then in the $y$-direction (see the arguments in $\mathrm{Op_y}$), with the subscript $a+1$ indicating the completion of one time step as the sum of both movements. This equation serves as the basis when convection plays a significant role in our numerical results. In the main text, we refer to $\Delta t_{Eu}$ as $\Delta t_{adv}$, since this time step is used exclusively during the advection integration. 

In summary, the approximations in equations (\ref{1st-FD}), (\ref{2nd-FD}), (\ref{BF-dif}) and (\ref{Ad_discre}), as well as the higher numerical methods for time integration from \texttt{DifferentialEquations.jl}, form the foundation of our upwind difference scheme. For more computational and numerical details, see \cite{salmeron2024dynamics} and the source code.

\section{Scaling and Nondimensionalization}
\label{appendix:c}

To relate the computational model to physical units, we define the mapping between the dimensionless variables used in simulation and their corresponding physical counterparts. We use the same variables, abbreviations, and symbols, as in the main text, unless stated otherwise.

The dimensionless formulation for the diffusion-relaxation step is obtained by scaling all spatial coordinates by a characteristic length scale $\xi_{N}$, commonly associated with the nematic coherence length  (although it might have a different meaning depending on the size of the phenomenon of interest), time by a characteristic time $t_{ch}$, and other relevant quantities by appropriate reference values ($T_{ch}$, $\rho_{ch}$). 

The dimensionless variables, denoted with a star superscript, are:
\begin{equation} \label{nondimensionalization}
\bm{r}^* = \frac{\bm{r}}{\xi_{N}}, \quad 
t^* = \frac{t}{t_{ch}}, \quad 
T^* = \frac{T}{T_{ch}}, \quad 
\rho^* = \frac{\rho}{\rho_{ch}}. \quad 
\end{equation}
While different non-dimensionalization variables are possible, this choice best matches the physical regime we aim to model and is consistent with the characteristic scales used in our numerical implementation. 

With these dimensionless parameters, Eqs. (\ref{CH_noflow}) and (\ref{GL_noflow}), together with Eqs. (\ref{Egrad}), (\ref{Sgrad}), (\ref{A_phi}), and (\ref{B_phi}), can be rewritten as:
\begin{equation}\label{CH_dimensionless}
\begin{split}
\frac{\partial\phi}{\partial t^*} = \tilde{D_\rho}  \frac{\partial^2}{\partial (r_l^*)^2}   
\left\{ \frac{1}{T^*} \left[  
  -\tilde{\kappa}' \frac{\partial}{\partial r_k^*}
 \!\!\left[\left( Q_{kl}
 + \frac{I_{kl}}{d}\right) 
  \frac{\partial \phi}{\partial r_l^*} \right]  - \tilde{\kappa}_E 
 \frac{\partial}{\partial r_k^*}\frac{\partial \phi}{\partial {r_k^*}} + \rho^* \tilde{\chi} (1 - 2\phi) 
 \right. \right. \\
\left. \left.
 -2 \rho^* \tilde{A_0}  \tilde{U}  \phi \left(
\frac{1}{6} \, Q_{ij}Q_{ji}
+ \frac{1}{3} \, Q_{ij}Q_{jk}Q_{ki}
- \frac{1}{4} \, (Q_{ij}Q_{ji})^2
\right) + \frac{\rho^* T^*}{2} \tilde{A_0} \,   \,Q_{ij}Q_{ji} \right]
\right. \\
\left.
+  \rho^*  \Big[  \tilde{C}_1 \ln \phi - \tilde{C}_2 \ln (1-\phi) + \tilde{C}_1 - \tilde{C}_2 \Big]  \right\} ,
\end{split}
\end{equation}
and
\begin{equation} \label{GL_dimensionless}
\begin{split}
\frac{\partial Q_{ij}}{\partial t^*} = - \frac{R}{T^*}  \left\{ 
\frac{1}{2} \tilde{\kappa}' \left[ \left(\frac{\partial \phi}{\partial r_i^*}\right) \left(\frac{\partial \phi}{\partial r_j^*}\right)
- \frac{I_{ij}}{d} \left(\frac{\partial \phi}{\partial r_l^*}\right)^2 \right]  - \tilde{K} \frac{\partial}{\partial r_k^*} \frac{\partial Q_{ij}}{\partial r_k^*}  
\right. \\
\left. - \rho^* \tilde{A_0}  \left[\tilde{U}  \phi^2\left( \frac{1}{3}Q_{ij} + Q_{ik}Q_{kj} -\left( Q_{ij} + \frac{I_{ij}}{d}\right) Q_{lk}Q_{kl}  \right) \,  - T^* \phi  \, Q_{ij} \right] \right\} ,
\end{split}
\end{equation}

where we denote the computational parameters with a tilde. Although they are not strictly dimensionless, they help bridge the physical parameters introduced in Section~\ref{Model} and their counterparts used in the simulations (Section~\ref{Results}). These parameters are defined as follows:
\begin{equation} \label{computational_parameters}
\begin{split}
\tilde{\kappa}' = \frac{\kappa'\, t_{ch}}{T_{ch} \, (\xi_N)^2}, \quad 
\tilde{K} = \frac{K \, t_{ch}}{T_{ch} \,(\xi_N)^2}, \quad 
\tilde{\kappa}_E = \frac{\kappa_E \, t_{ch}}{T_{ch} \,(\xi_N)^2}, \quad
\tilde{C}_1 = \rho_{ch} \, C_1 \, t_{ch} ,
\\ 
\tilde{A_0} = \rho_{ch}  A_0 \, t_{ch}, \quad 
\tilde{U} = \frac{U}{T_{ch} }, \quad 
\tilde{D_\rho} = \frac{D_\rho}{(\xi_N)^2}, \quad
\tilde{\chi} =  \frac{\rho_{ch} \, \chi \,  t_{ch}}{T_{ch}},\quad
\tilde{C}_2 = \rho_{ch} \, C_2 \, t_{ch} .
\end{split}
\end{equation}
Eqs. (\ref{CH_dimensionless}) and (\ref{GL_dimensionless}) are the equations solved in the simulations. However, for the sake of simplicity, we omit the tilde and star symbols in the main text. Note that the parameter $R$ remains unchanged. To recover physical values, we use Eqs. (\ref{nondimensionalization}) and (\ref{computational_parameters}). For instance, the physical temperature is computed as $T = T_{ch} \cdot T^*$.

To illustrate our choices and their agreement with physical quantities, we compare our computational and dimensionless parameters with values reported in the literature. Because some physical parameters are not readily available or vary between sources, especially since they are often material-dependent, we adopt a pragmatic approach by focusing on orders of magnitude. This keeps the model grounded while allowing some flexibility to apply it in different scenarios—for example, when all physical values are known from the same experimental dataset. We start from the most consistently reported quantities and use them to estimate the rest, as they constrain the remaining values. All physical quantities are understood to be expressed in SI (\emph{Système International d’Unités}) units to ensure consistency and compatibility across parameters; however, units are not explicitly indicated for the sake of conciseness.

We begin by using the simplest parameters and noting that $\rho_{ch} \sim 10^3$ (typical liquid density) and $T_{ch} \sim 10^2$ (room temperature), which implies $T^* = 1$ and $\rho^* = 1$ (as in the main text). These values are held constant throughout this paper because the process is considered isothermal, and total density variations are neglected (i.e., the mixture of the two liquids is assumed incompressible).

Next, the first coefficient in the LdG free energy (for a pure LC, $\phi\approx 1$, as chosen for simplicity in the simulation results) is reported as $A_{lit} \sim 10^5$ \cite{Sengupta2013}. In our case, this corresponds to the product:
\begin{equation}\label{A_lit}
    A_{lit} = \rho A_0 \, T \sim 10^5 \sim 10^3 \cdot 10^2 \cdot A_0 ,
\end{equation}
and therefore our physical $A_0 \sim 10^0$. Since our computational $\tilde{U} \sim 10$, we recover $U = T_{ch} \cdot \tilde{U} \sim 10^2 \cdot 10 \sim 10^3$, which is consistent with reported values for the second LdG coefficient \cite{Sengupta2013}, $B_{lit} \sim 10^6$, expressed here as:
\begin{equation}\label{B_lit}
    B_{lit} = \rho \, A_0 \, U \sim 10^3 \cdot 10^0 \cdot 10^3 \sim 10^6.
\end{equation}
Notice that in some works, the first LdG coefficient takes the form $\rho A_0 (T - T_{ref})$, which we do not use here, as we assume our model operates near $T_{ref}$. In those instances, the reference temperature is typically $T_{ref} \sim 10^2$, implying $T - T_{ref} \sim 10^0$. This would shift $A_0$ by two orders of magnitude, potentially leading to a significant reduction in the physical time (see below), or simply requiring different overall choices for the dimensionless parameters.

Furthermore, considering that our computational $\tilde{A_0} \sim 10^0$, we use its definition to estimate the characteristic time as:
\begin{equation} \label{t_ch_estimate}
    t_{ch} = \frac{\tilde{A_0}}{\rho_{ch} \, A_0} \sim 10^{-3}.
\end{equation}
Using this value, we approximate the total simulation duration in Section \ref{Equilibrium_integration}, given that the dimensionless simulation time is $t^* \sim 10^5$:
\begin{equation}\label{physical_time}
    t = t_{ch} \cdot t^* \sim 10^{-3} 10^{5} \sim 10^2 ,   
\end{equation}
which suggests that $t$ is on the order of minutes to tens of minutes. Although the exact physical timescale of the experiment remains, to the best of our knowledge, uncertain, we consider this a reasonable approximation, especially since the simulation is initialized with an ansatz that accelerates the director vector relaxation.

Continuing, we approximate the length scale $\xi_N$ as
\begin{equation}\label{xi_N_estimate}
    \xi_N = \sqrt{\frac{K \, t_{ch}}{T_{ch} \,\tilde{K}} } \sim \sqrt{\frac{10^{-10} 10^{-3}}{10^{2}}} \sim 10^{-7} ,
\end{equation}
where we take $\tilde{K}\sim10^0$ from the main text, and $K \sim10^{-10}$, a typical value for the experimental elastic constant of lyotropic LCs in solution \cite{zhou2018recent}. The resulting estimate $\xi_N \sim 10^{-7}$ is also consistent with previous work, especially with the simulated droplet size being $r_d = 15dx = 15 \xi_N \sim 10^{-6}$, placing their size in the micron range, in agreement with what is observed experimentally (see the scale bars in Fig. \ref{Lav_exp}).

Analogously, once the range of $\xi_N$ is fixed, it constrains the physical values of $\kappa_E \sim 10^{-10}$ and $\kappa' \sim 10^{-10}$, given the similarity between their computational definitions and that of the elastic constant. The diffuse interface constant tends to align well with reported values.

We recover \(D_{\rho} = \tilde{D_\rho} \, (\xi_N)^2 \sim 10^{-2} 10^{-15} \sim 10^{-17}\), and \(\chi = T_{ch} \, \tilde{\chi} /\rho_{ch} \, t_{ch} \sim 10^{2}\). To compare, we express the FH parameter in the literature as \(\chi_{\text{lit}} = \rho \, \chi \sim 10^{5}\). The anchoring and FH parameters are among the most challenging to estimate and often show inconsistency in the literature, particularly for specific materials or the units in which they are expressed. As a result, they require computational tuning, especially the anchoring constant, which is a relatively recent and unmeasured parameter. However, we argue that their physical orders of magnitude here remain acceptable.

Finally, we estimate the values of the mixing entropy constants as \(C_1 = \tilde{C}_1 / (\rho_{ch} \, t_{ch}) \sim 10^0\), which is also the order of magnitude of \(C_2\). Notably, as indicated in the discussion of the main text, this parameter represents Boltzmann’s constant (\(k_B\)) divided by the mass of a single molecule of one of the mixture components. From this, we infer the mass of this molecule as \(m_1 = k_B / C_1 \sim 10^{-23}\), which corresponds to the size of a small polymer. For comparison, the mass of a DSCG molecule is on the order of \(\sim 10^{-24}\), according to its chemical formula. We therefore assert that our computational model provides a valid theoretical prediction of the same physics observed in a DSCG--water tactoid experiment, albeit with different concentration profiles. Since our \(C_1\) and \(C_2\) are taken to be equal (and relatively small), they imply a slightly larger molecular size and consequently shift the equilibrium concentration ($\phi_e$), which might impact the total time of the relaxation-diffusion integration.

For completeness, we also show the computational mapping of the advection step to illustrate that it involves a different characteristic timescale. The associated dimensionless parameters are:
\begin{equation} \label{nondimensionalization_advection}
t_{adv}^* = \frac{t}{t_{ca}}, \quad 
v_x^* = \frac{v_x t_{ca}}{\xi_N}, \quad 
\end{equation}
which yields the following dimensionless advection equation
\begin{equation}
 \frac{\partial \phi}{\partial t_{adv}^*} = -v_x^* \frac{\partial \phi}{\partial x^*}    ,
\end{equation}
here we use $\phi$, although it could readily be replaced by $\bm{Q}$; $x^*$ denotes the dimensionless Cartesian coordinate (see Eq.~\ref{nondimensionalization}).

In this advection equation, we only consider the $x$-component of the velocity field ($v_x$)  because our flows are unidirectional, though incorporating the other directions is straightforward. We also define the velocity using the same length scale ($\xi_N$), since we employ the same discretized structural mesh as in the diffusion-relaxation step, making it a convenient choice (this could change in more complex models involving adaptive or differently sized meshes for each step).

Both the velocity and the time in the advective step ($t_{adv}^*$) are  parametrized by $t_{ca}$, the characteristic time of advection. We consider $t_{ca}$ an independent degree of freedom, which constrains the physical value of the velocity to be recovered (or vice versa). For instance, it could be linked to an experimentally observable timescale, either directly measured for a given phenomenon or inferred from a known flow velocity. 

To estimate $t_{adv}^*$, we recall from the main text that the computational advection time step ($\Delta t_{adv}$) is inversely proportional to the diffusion-relaxation time step ($\Delta t_{dr}$). Thus, as a working hypothesis in the semi-Couette flow configuration, we consider  $t_{ca} \sim 10^{-1} \, t_{ch} \sim 10^{-4}$ (based on the ratio of the two computational time steps), which yields $v_x = v_x^* * \xi_N / t_{ca} \sim 10^{-2} 10^{-7} / 10^{-4} \sim 10^{-5}$ (given that $v_x^* \sim 10^{-2}$ in the main text), a reasonable value  compared to microfluidic channel experiments \cite{Tadej2019Sculpting}. However, this estimate should be treated with caution, as it is not guaranteed that the two processes (diffusion-relaxation and advection) follow a direct correlation.

Overall, we demonstrate that this systematic computational-dimensionless-physical mapping provides a robust foundation for parameter tuning and selection in future LC simulations, and even in more complex systems of DEs aimed at reproducing experimental data by explicitly correlating physical parameters with those employed in computational modeling.

\section{Integration Algorithm}
\label{Appex:Flowchart}

\begin{figure}[hbt!]
    \centering

        \includegraphics[scale=0.7]{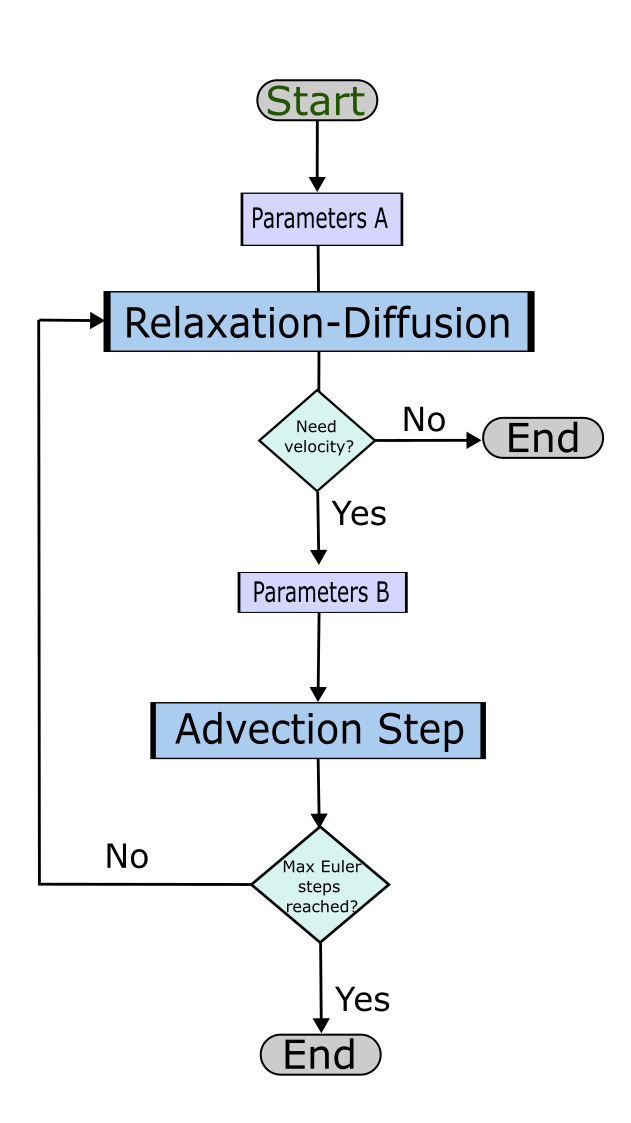}

    \caption{Basic Integration Steps for the Advection–Diffusion–Relaxation System}
    \label{Flowchart}
\end{figure}

The flow diagram in Figure~\ref{Flowchart} shows the basic steps of our advection-diffusion-relaxation algorithm.  We use the same variables, abbreviations, and symbols, as in the main text, unless stated otherwise. Parameter set A includes: 1) ICs and BCs, 2) Physical parameters (e.g., $T$, $U$, etc.), 3) Relaxation–diffusion time settings (e.g., $t_f$, integrator, $\Delta t$), 4) Mesh ($N_x$, $N_y$, $\Delta x$, $\Delta y$), 5) Centered-difference operators, and 6) Cached terms for intermediate calculations. Parameter set B comprises: 1) Velocity field, 2) Advection time settings (number of Euler steps, $\Delta t_{adv}$), and 3) Upwind scheme operators and caches. 

\bibliography{bibliography}

\end{document}